\documentclass[twocolumn]{aastex631}

\usepackage{amsmath}
\usepackage{amssymb}
\usepackage{wasysym}
\usepackage{needspace}
\usepackage{xurl}

\newcommand{\Mch}{M_{\text{ch}}}
\newcommand{\Msun}{M_{\astrosun}}

\newcommand{\CURN}{{\text{CURN}}}

\newcommand{\gwecc}{\texttt{GWecc.jl}}

\begin{document}

\title{The NANOGrav 12.5-year data set: A computationally efficient eccentric binary search pipeline and constraints on an eccentric supermassive binary candidate in 3C~66B}

\correspondingauthor{Lankeswar Dey}
\email{lankeswar.dey@nanograv.org}

\correspondingauthor{Abhimanyu Susobhanan}
\email{abhimanyu.susobhanan@nanograv.org}

\author[0000-0001-5134-3925]{Gabriella Agazie}
\affiliation{Center for Gravitation, Cosmology and Astrophysics, Department of Physics, University of Wisconsin-Milwaukee,\\ P.O. Box 413, Milwaukee, WI 53201, USA}
\author{Zaven Arzoumanian}
\affiliation{X-Ray Astrophysics Laboratory, NASA Goddard Space Flight Center, Code 662, Greenbelt, MD 20771, USA}
\author[0000-0003-2745-753X]{Paul T. Baker}
\affiliation{Department of Physics and Astronomy, Widener University, One University Place, Chester, PA 19013, USA}
\author[0000-0003-0909-5563]{Bence Bécsy}
\affiliation{Department of Physics, Oregon State University, Corvallis, OR 97331, USA}
\author[0000-0002-2183-1087]{Laura Blecha}
\affiliation{Physics Department, University of Florida, Gainesville, FL 32611, USA}
\author[0000-0003-4046-884X]{Harsha Blumer}
\affiliation{Department of Physics and Astronomy, West Virginia University, P.O. Box 6315, Morgantown, WV 26506, USA}
\affiliation{Center for Gravitational Waves and Cosmology, West Virginia University, Chestnut Ridge Research Building, Morgantown, WV 26505, USA}
\author[0000-0001-6341-7178]{Adam Brazier}
\affiliation{Cornell Center for Astrophysics and Planetary Science and Department of Astronomy, Cornell University, Ithaca, NY 14853, USA}
\affiliation{Cornell Center for Advanced Computing, Cornell University, Ithaca, NY 14853, USA}
\author[0000-0003-3053-6538]{Paul R. Brook}
\affiliation{Institute for Gravitational Wave Astronomy and School of Physics and Astronomy, University of Birmingham, Edgbaston, Birmingham B15 2TT, UK}
\author[0000-0003-4052-7838]{Sarah Burke-Spolaor}
\altaffiliation{Sloan Fellow}
\affiliation{Department of Physics and Astronomy, West Virginia University, P.O. Box 6315, Morgantown, WV 26506, USA}
\affiliation{Center for Gravitational Waves and Cosmology, West Virginia University, Chestnut Ridge Research Building, Morgantown, WV 26505, USA}
\author[0000-0002-5557-4007]{J. Andrew Casey-Clyde}
\affiliation{Department of Physics, University of Connecticut, 196 Auditorium Road, U-3046, Storrs, CT 06269-3046, USA}
\author[0000-0003-3579-2522]{Maria Charisi}
\affiliation{Department of Physics and Astronomy, Vanderbilt University, 2301 Vanderbilt Place, Nashville, TN 37235, USA}
\author[0000-0002-2878-1502]{Shami Chatterjee}
\affiliation{Cornell Center for Astrophysics and Planetary Science and Department of Astronomy, Cornell University, Ithaca, NY 14853, USA}
\author{Belinda D. Cheeseboro}
\affiliation{Department of Physics and Astronomy, West Virginia University, P.O. Box 6315, Morgantown, WV 26506, USA}
\affiliation{Center for Gravitational Waves and Cosmology, West Virginia University, Chestnut Ridge Research Building, Morgantown, WV 26505, USA}
\author[0000-0001-7587-5483]{Tyler Cohen}
\affiliation{Department of Physics, New Mexico Institute of Mining and Technology, 801 Leroy Place, Socorro, NM 87801, USA}
\author[0000-0002-4049-1882]{James M. Cordes}
\affiliation{Cornell Center for Astrophysics and Planetary Science and Department of Astronomy, Cornell University, Ithaca, NY 14853, USA}
\author[0000-0002-7435-0869]{Neil J. Cornish}
\affiliation{Department of Physics, Montana State University, Bozeman, MT 59717, USA}
\author[0000-0002-2578-0360]{Fronefield Crawford}
\affiliation{Department of Physics and Astronomy, Franklin \& Marshall College, P.O. Box 3003, Lancaster, PA 17604, USA}
\author[0000-0002-6039-692X]{H. Thankful Cromartie}
\altaffiliation{NASA Hubble Fellowship: Einstein Postdoctoral Fellow}
\affiliation{Cornell Center for Astrophysics and Planetary Science and Department of Astronomy, Cornell University, Ithaca, NY 14853, USA}
\author[0000-0002-2185-1790]{Megan E. DeCesar}
\affiliation{George Mason University, resident at the Naval Research Laboratory, Washington, DC 20375, USA}
\author[0000-0002-6664-965X]{Paul B. Demorest}
\affiliation{National Radio Astronomy Observatory, 1003 Lopezville Rd., Socorro, NM 87801, USA}
\author[0000-0002-2554-0674]{Lankeswar Dey}
\affiliation{Department of Physics and Astronomy, West Virginia University, P.O. Box 6315, Morgantown, WV 26506, USA}
\affiliation{Center for Gravitational Waves and Cosmology, West Virginia University, Chestnut Ridge Research Building, Morgantown, WV 26505, USA}
\author[0000-0001-8885-6388]{Timothy Dolch}
\affiliation{Department of Physics, Hillsdale College, 33 E. College Street, Hillsdale, MI 49242, USA}
\affiliation{Eureka Scientific, 2452 Delmer Street, Suite 100, Oakland, CA 94602-3017, USA}
\author{Justin A. Ellis}
\altaffiliation{Infinia ML, 202 Rigsbee Avenue, Durham NC, 27701, USA}
\author[0000-0002-2223-1235]{Robert D. Ferdman}
\affiliation{School of Chemistry, University of East Anglia, Norwich, NR4 7TJ, United Kingdom}
\author[0000-0001-7828-7708]{Elizabeth C. Ferrara}
\affiliation{Department of Astronomy, University of Maryland, College Park, MD 20742, USA}
\affiliation{Center for Research and Exploration in Space Science and Technology, NASA/GSFC, Greenbelt, MD 20771}
\affiliation{NASA Goddard Space Flight Center, Greenbelt, MD 20771, USA}
\author[0000-0001-5645-5336]{William Fiore}
\affiliation{Department of Physics and Astronomy, West Virginia University, P.O. Box 6315, Morgantown, WV 26506, USA}
\affiliation{Center for Gravitational Waves and Cosmology, West Virginia University, Chestnut Ridge Research Building, Morgantown, WV 26505, USA}
\author[0000-0001-8384-5049]{Emmanuel Fonseca}
\affiliation{Department of Physics and Astronomy, West Virginia University, P.O. Box 6315, Morgantown, WV 26506, USA}
\affiliation{Center for Gravitational Waves and Cosmology, West Virginia University, Chestnut Ridge Research Building, Morgantown, WV 26505, USA}
\author[0000-0001-7624-4616]{Gabriel E. Freedman}
\affiliation{Center for Gravitation, Cosmology and Astrophysics, Department of Physics, University of Wisconsin-Milwaukee,\\ P.O. Box 413, Milwaukee, WI 53201, USA}
\author[0000-0001-6166-9646]{Nate Garver-Daniels}
\affiliation{Department of Physics and Astronomy, West Virginia University, P.O. Box 6315, Morgantown, WV 26506, USA}
\affiliation{Center for Gravitational Waves and Cosmology, West Virginia University, Chestnut Ridge Research Building, Morgantown, WV 26505, USA}
\author[0000-0001-8158-683X]{Peter A. Gentile}
\affiliation{Department of Physics and Astronomy, West Virginia University, P.O. Box 6315, Morgantown, WV 26506, USA}
\affiliation{Center for Gravitational Waves and Cosmology, West Virginia University, Chestnut Ridge Research Building, Morgantown, WV 26505, USA}
\author[0000-0003-4090-9780]{Joseph Glaser}
\affiliation{Department of Physics and Astronomy, West Virginia University, P.O. Box 6315, Morgantown, WV 26506, USA}
\affiliation{Center for Gravitational Waves and Cosmology, West Virginia University, Chestnut Ridge Research Building, Morgantown, WV 26505, USA}
\author[0000-0003-1884-348X]{Deborah C. Good}
\affiliation{Department of Physics, University of Connecticut, 196 Auditorium Road, U-3046, Storrs, CT 06269-3046, USA}
\affiliation{Center for Computational Astrophysics, Flatiron Institute, 162 5th Avenue, New York, NY 10010, USA}
\author[0000-0003-4274-4369]{Achamveedu Gopakumar}
\affiliation{Department of Astronomy and Astrophysics, Tata Institute of Fundamental Research, Dr. Homi Bhabha Road, Mumbai, Maharashtra 400005, India}
\author[0000-0002-1146-0198]{Kayhan Gültekin}
\affiliation{Department of Astronomy and Astrophysics, University of Michigan, Ann Arbor, MI 48109, USA}
\author[0000-0003-2742-3321]{Jeffrey S. Hazboun}
\affiliation{Department of Physics, Oregon State University, Corvallis, OR 97331, USA}
\author[0000-0003-1082-2342]{Ross J. Jennings}
\altaffiliation{NANOGrav Physics Frontiers Center Postdoctoral Fellow}
\affiliation{Department of Physics and Astronomy, West Virginia University, P.O. Box 6315, Morgantown, WV 26506, USA}
\affiliation{Center for Gravitational Waves and Cosmology, West Virginia University, Chestnut Ridge Research Building, Morgantown, WV 26505, USA}
\author[0000-0002-7445-8423]{Aaron D. Johnson}
\affiliation{Center for Gravitation, Cosmology and Astrophysics, Department of Physics, University of Wisconsin-Milwaukee,\\ P.O. Box 413, Milwaukee, WI 53201, USA}
\affiliation{Division of Physics, Mathematics, and Astronomy, California Institute of Technology, Pasadena, CA 91125, USA}
\author[0000-0001-6607-3710]{Megan L. Jones}
\affiliation{Center for Gravitation, Cosmology and Astrophysics, Department of Physics, University of Wisconsin-Milwaukee,\\ P.O. Box 413, Milwaukee, WI 53201, USA}
\author[0000-0002-3654-980X]{Andrew R. Kaiser}
\affiliation{Department of Physics and Astronomy, West Virginia University, P.O. Box 6315, Morgantown, WV 26506, USA}
\affiliation{Center for Gravitational Waves and Cosmology, West Virginia University, Chestnut Ridge Research Building, Morgantown, WV 26505, USA}
\author[0000-0001-6295-2881]{David L. Kaplan}
\affiliation{Center for Gravitation, Cosmology and Astrophysics, Department of Physics, University of Wisconsin-Milwaukee,\\ P.O. Box 413, Milwaukee, WI 53201, USA}
\author[0000-0002-6625-6450]{Luke Zoltan Kelley}
\affiliation{Department of Astronomy, University of California, Berkeley, 501 Campbell Hall \#3411, Berkeley, CA 94720, USA}
\author[0000-0003-0123-7600]{Joey S. Key}
\affiliation{University of Washington Bothell, 18115 Campus Way NE, Bothell, WA 98011, USA}
\author[0000-0002-9197-7604]{Nima Laal}
\affiliation{Department of Physics, Oregon State University, Corvallis, OR 97331, USA}
\author[0000-0003-0721-651X]{Michael T. Lam}
\affiliation{SETI Institute, 339 N Bernardo Ave Suite 200, Mountain View, CA 94043, USA}
\affiliation{School of Physics and Astronomy, Rochester Institute of Technology, Rochester, NY 14623, USA}
\affiliation{Laboratory for Multiwavelength Astrophysics, Rochester Institute of Technology, Rochester, NY 14623, USA}
\author[0000-0003-1096-4156]{William G. Lamb}
\affiliation{Department of Physics and Astronomy, Vanderbilt University, 2301 Vanderbilt Place, Nashville, TN 37235, USA}
\author{T. Joseph W. Lazio}
\affiliation{Jet Propulsion Laboratory, California Institute of Technology, 4800 Oak Grove Drive, Pasadena, CA 91109, USA}
\author[0000-0003-0771-6581]{Natalia Lewandowska}
\affiliation{Department of Physics, State University of New York at Oswego, Oswego, NY 13126, USA}
\author[0000-0001-5766-4287]{Tingting Liu}
\affiliation{Department of Physics and Astronomy, West Virginia University, P.O. Box 6315, Morgantown, WV 26506, USA}
\affiliation{Center for Gravitational Waves and Cosmology, West Virginia University, Chestnut Ridge Research Building, Morgantown, WV 26505, USA}
\author[0000-0003-1301-966X]{Duncan R. Lorimer}
\affiliation{Department of Physics and Astronomy, West Virginia University, P.O. Box 6315, Morgantown, WV 26506, USA}
\affiliation{Center for Gravitational Waves and Cosmology, West Virginia University, Chestnut Ridge Research Building, Morgantown, WV 26505, USA}
\author[0000-0001-5373-5914]{Jing Luo}
\altaffiliation{Deceased}
\affiliation{Department of Astronomy \& Astrophysics, University of Toronto, 50 Saint George Street, Toronto, ON M5S 3H4, Canada}
\author[0000-0001-5229-7430]{Ryan S. Lynch}
\affiliation{Green Bank Observatory, P.O. Box 2, Green Bank, WV 24944, USA}
\author[0000-0002-4430-102X]{Chung-Pei Ma}
\affiliation{Department of Astronomy, University of California, Berkeley, 501 Campbell Hall \#3411, Berkeley, CA 94720, USA}
\affiliation{Department of Physics, University of California, Berkeley, CA 94720, USA}
\author[0000-0003-2285-0404]{Dustin R. Madison}
\affiliation{Department of Physics, University of the Pacific, 3601 Pacific Avenue, Stockton, CA 95211, USA}
\author[0000-0001-5481-7559]{Alexander McEwen}
\affiliation{Center for Gravitation, Cosmology and Astrophysics, Department of Physics, University of Wisconsin-Milwaukee,\\ P.O. Box 413, Milwaukee, WI 53201, USA}
\author[0000-0002-2885-8485]{James W. McKee}
\affiliation{E.A. Milne Centre for Astrophysics, University of Hull, Cottingham Road, Kingston-upon-Hull, HU6 7RX, UK}
\affiliation{Centre of Excellence for Data Science, Artificial Intelligence and Modelling (DAIM), University of Hull, Cottingham Road, Kingston-upon-Hull, HU6 7RX, UK}
\author[0000-0001-7697-7422]{Maura A. McLaughlin}
\affiliation{Department of Physics and Astronomy, West Virginia University, P.O. Box 6315, Morgantown, WV 26506, USA}
\affiliation{Center for Gravitational Waves and Cosmology, West Virginia University, Chestnut Ridge Research Building, Morgantown, WV 26505, USA}
\author[0000-0002-2689-0190]{Patrick M. Meyers}
\affiliation{Division of Physics, Mathematics, and Astronomy, California Institute of Technology, Pasadena, CA 91125, USA}
\author[0000-0002-4307-1322]{Chiara M. F. Mingarelli}
\affiliation{Department of Physics, Yale University, New Haven, CT 06520, USA}
\author[0000-0003-2898-5844]{Andrea Mitridate}
\affiliation{Deutsches Elektronen-Synchrotron DESY, Notkestr. 85, 22607 Hamburg, Germany}
\author[0000-0002-3616-5160]{Cherry Ng}
\affiliation{Dunlap Institute for Astronomy and Astrophysics, University of Toronto, 50 St. George St., Toronto, ON M5S 3H4, Canada}
\author[0000-0002-6709-2566]{David J. Nice}
\affiliation{Department of Physics, Lafayette College, Easton, PA 18042, USA}
\author[0000-0002-4941-5333]{Stella Koch Ocker}
\affiliation{Division of Physics, Mathematics, and Astronomy, California Institute of Technology, Pasadena, CA 91125, USA}
\affiliation{The Observatories of the Carnegie Institution for Science, Pasadena, CA 91101, USA}
\author[0000-0002-2027-3714]{Ken D. Olum}
\affiliation{Institute of Cosmology, Department of Physics and Astronomy, Tufts University, Medford, MA 02155, USA}
\author[0000-0001-5465-2889]{Timothy T. Pennucci}
\affiliation{Institute of Physics and Astronomy, E\"{o}tv\"{o}s Lor\'{a}nd University, P\'{a}zm\'{a}ny P. s. 1/A, 1117 Budapest, Hungary}
\author[0000-0002-8826-1285]{Nihan S. Pol}
\affiliation{Department of Physics and Astronomy, Vanderbilt University, 2301 Vanderbilt Place, Nashville, TN 37235, USA}
\author[0000-0002-2074-4360]{Henri A. Radovan}
\affiliation{Department of Physics, University of Puerto Rico, Mayag\"{u}ez, PR 00681, USA}
\author[0000-0001-5799-9714]{Scott M. Ransom}
\affiliation{National Radio Astronomy Observatory, 520 Edgemont Road, Charlottesville, VA 22903, USA}
\author[0000-0002-5297-5278]{Paul S. Ray}
\affiliation{Space Science Division, Naval Research Laboratory, Washington, DC 20375-5352, USA}
\author[0000-0003-4915-3246]{Joseph D. Romano}
\affiliation{Department of Physics, Texas Tech University, Box 41051, Lubbock, TX 79409, USA}
\author[0009-0006-5476-3603]{Shashwat C. Sardesai}
\affiliation{Center for Gravitation, Cosmology and Astrophysics, Department of Physics, University of Wisconsin-Milwaukee,\\ P.O. Box 413, Milwaukee, WI 53201, USA}
\author[0000-0003-2807-6472]{Kai Schmitz}
\affiliation{Institute for Theoretical Physics, University of Münster, 48149 Münster, Germany}
\author[0000-0002-7778-2990]{Xavier Siemens}
\affiliation{Department of Physics, Oregon State University, Corvallis, OR 97331, USA}
\affiliation{Center for Gravitation, Cosmology and Astrophysics, Department of Physics, University of Wisconsin-Milwaukee,\\ P.O. Box 413, Milwaukee, WI 53201, USA}
\author[0000-0003-1407-6607]{Joseph Simon}
\altaffiliation{NSF Astronomy and Astrophysics Postdoctoral Fellow}
\affiliation{Department of Astrophysical and Planetary Sciences, University of Colorado, Boulder, CO 80309, USA}
\author[0000-0002-1530-9778]{Magdalena S. Siwek}
\affiliation{Center for Astrophysics, Harvard University, 60 Garden St, Cambridge, MA 02138, USA}
\author[0000-0002-5176-2924]{Sophia V. Sosa Fiscella}
\affiliation{School of Physics and Astronomy, Rochester Institute of Technology, Rochester, NY 14623, USA}
\affiliation{Laboratory for Multiwavelength Astrophysics, Rochester Institute of Technology, Rochester, NY 14623, USA}
\author[0000-0002-6730-3298]{Renée Spiewak}
\affiliation{Jodrell Bank Centre for Astrophysics, University of Manchester, Manchester, M13 9PL, United Kingdom}
\author[0000-0001-9784-8670]{Ingrid H. Stairs}
\affiliation{Department of Physics and Astronomy, University of British Columbia, 6224 Agricultural Road, Vancouver, BC V6T 1Z1, Canada}
\author[0000-0002-1797-3277]{Daniel R. Stinebring}
\affiliation{Department of Physics and Astronomy, Oberlin College, Oberlin, OH 44074, USA}
\author[0000-0002-7261-594X]{Kevin Stovall}
\affiliation{National Radio Astronomy Observatory, 1003 Lopezville Rd., Socorro, NM 87801, USA}
\author[0000-0002-2820-0931]{Abhimanyu Susobhanan}
\affiliation{Center for Gravitation, Cosmology and Astrophysics, Department of Physics, University of Wisconsin-Milwaukee,\\ P.O. Box 413, Milwaukee, WI 53201, USA}
\author[0000-0002-1075-3837]{Joseph K. Swiggum}
\altaffiliation{NANOGrav Physics Frontiers Center Postdoctoral Fellow}
\affiliation{Department of Physics, Lafayette College, Easton, PA 18042, USA}
\author[0000-0003-0264-1453]{Stephen R. Taylor}
\affiliation{Department of Physics and Astronomy, Vanderbilt University, 2301 Vanderbilt Place, Nashville, TN 37235, USA}
\author[0000-0002-2451-7288]{Jacob E. Turner}
\affiliation{Green Bank Observatory, P.O. Box 2, Green Bank, WV 24944, USA}
\author[0000-0001-8800-0192]{Caner Unal}
\affiliation{Department of Physics, Middle East Technical University, 06531 Ankara, Turkey}
\affiliation{Department of Physics, Ben-Gurion University of the Negev, Be'er Sheva 84105, Israel}
\affiliation{Feza Gursey Institute, Bogazici University, Kandilli, 34684, Istanbul, Turkey}
\author[0000-0002-4162-0033]{Michele Vallisneri}
\affiliation{Jet Propulsion Laboratory, California Institute of Technology, 4800 Oak Grove Drive, Pasadena, CA 91109, USA}
\affiliation{Division of Physics, Mathematics, and Astronomy, California Institute of Technology, Pasadena, CA 91125, USA}
\author[0000-0003-4700-9072]{Sarah J. Vigeland}
\affiliation{Center for Gravitation, Cosmology and Astrophysics, Department of Physics, University of Wisconsin-Milwaukee,\\ P.O. Box 413, Milwaukee, WI 53201, USA}
\author[0000-0002-6020-9274]{Caitlin A. Witt}
\affiliation{Center for Interdisciplinary Exploration and Research in Astrophysics (CIERA), Northwestern University, Evanston, IL 60208, USA}
\affiliation{Adler Planetarium, 1300 S. DuSable Lake Shore Dr., Chicago, IL 60605, USA}
\author[0000-0002-0883-0688]{Olivia Young}
\affiliation{School of Physics and Astronomy, Rochester Institute of Technology, Rochester, NY 14623, USA}
\affiliation{Laboratory for Multiwavelength Astrophysics, Rochester Institute of Technology, Rochester, NY 14623, USA}

\collaboration{90}{(The NANOGrav Collaboration)}








\begin{abstract}
The radio galaxy 3C~66B has been hypothesized to host a supermassive black hole binary (SMBHB) at its center based on electromagnetic observations. Its apparent 1.05-year period and low redshift ($\sim0.02$) make it an interesting testbed to search for low-frequency gravitational waves (GWs) using Pulsar Timing Array (PTA) experiments. This source has been subjected to multiple searches for continuous GWs from a circular SMBHB, resulting in progressively more stringent constraints on its GW amplitude and chirp mass. 
In this paper, we develop a pipeline for performing Bayesian targeted searches for eccentric SMBHBs in PTA data sets, and test its efficacy by applying it on simulated data sets with varying injected signal strengths. 
We also search for a realistic eccentric SMBHB source in 3C~66B using the NANOGrav 12.5-year data set employing PTA signal models containing Earth term-only as well as Earth+Pulsar term contributions using this pipeline.
Due to limitations in our PTA signal model, we get meaningful results only when the initial eccentricity $e_0<0.5$ and the symmetric mass ratio $\eta>0.1$.
We find no evidence for an eccentric SMBHB signal in our data, and therefore place 95\% upper limits on the PTA signal amplitude of $88.1\pm3.7$~ns for the Earth term-only and $81.74\pm0.86$~ns for the Earth+Pulsar term searches for $e_0<0.5$ and $\eta>0.1$.
Similar 95\%  upper limits on the chirp mass are $(1.98\pm0.05)\times10^9\,\Msun$ and $(1.81\pm0.01)\times10^9\,\Msun$.  
These upper limits, while less stringent than those calculated from a circular binary search in the NANOGrav 12.5-year data set, are consistent with the SMBHB model of 3C~66B developed from electromagnetic observations.
\end{abstract}



\section{Introduction} \label{sec:intro}

Routine detections of gravitational waves (GWs) from coalescing stellar-mass compact object binaries by the ground-based LIGO-Virgo-KAGRA observatories have opened a new window to our Universe \cite[e.g.][]{AbbottAbbott+2021}. 
While the ground-based observatories are sensitive to GW signals in the tens of hertz to kilohertz frequency range, Pulsar Timing Array \cite[PTA:][]{Sazhin1978,Detweiler1979,FosterBacker1990} 
experiments are sensitive to GWs in the nanohertz (nHz) frequency range.
PTAs achieve this by routinely monitoring ensembles of millisecond pulsars using some of the world's most sensitive radio telescopes to use them as accurate celestial clocks \citep{HobbsGuo+2019}.
There are six PTA collaborations active at present: 
the North American Nanohertz Observatory for Gravitational Waves \cite[NANOGrav:][]{DemorestFerdman+2013},
the European Pulsar Timing Array \cite[EPTA:][]{DesvignesCaballero+2016},
the Parkes Pulsar Timing Array \cite[PPTA:][]{ManchesterHobbs+2013},
the Indian Pulsar Timing Array \cite[InPTA:][]{TarafdarNobleson+2022},
the MeerKAT Pulsar Timing Array \cite[MPTA:][]{MilesShannon+2023}, and
the Chinese Pulsar Timing Array \cite[CPTA:][]{XuChen+2023}.
The International Pulsar Timing Array \cite[IPTA:][]{VerbiestLentati+2016} consortium combines data and resources from a subset of these collaborations and fosters scientific discourse to advance the prospects of nHz GW detection and post-discovery science.

Recently, evidence for the presence of a low-frequency stochastic GW background (GWB) in their respective PTA data sets was reported independently by NANOGrav \citep{NG2023_15yr_GWB}, EPTA+InPTA \citep{AntoniadisArumugam+2023c}, PPTA \citep{ReardonZic+2023}, and CPTA \citep{XuChen+2023}.
This signal manifests as a common-spectrum red noise process with a cross-pulsar spatial correlation that is consistent with the expected Hellings–Downs overlap reduction function \citep{HellingsDowns1983}.
These exciting results have inaugurated the era of nHz GW astronomy and the upcoming IPTA Data Release 3 is expected to strengthen the prospects of PTA science further.
The source(s) of this low-frequency GWB remain(s) inconclusive, and the observed GWB is consistent with many different physical processes in our universe, e.g., a population of supermassive black hole binaries (SMBHBs) emitting GWs, cosmic inflation, 
first-order phase transitions, and cosmic strings \citep[e.g.][]{NG2023_15yr_astro,AfzalAgazie+2023,EPTA_DR2_2023_5}.
Nevertheless, an ensemble of inspiralling SMBHBs present in active galactic nuclei (AGNs) is believed to be the most prominent source of the nHz GWB \citep{Burke-SpolaorTaylor+2019}.

SMBHBs are expected to form via galaxy mergers.
Observations suggest that nearly every galaxy contains a supermassive black hole (SMBH) at its center \citep{RichstoneAjhar+1998}.
When two galaxies merge, the pair of SMBHs interact with the broader merger environment during its early evolution; dynamical friction with the surrounding stars and gas helps the SMBH pair to efficiently sink towards the center of the merger remnant, eventually forming a bound SMBHB system \citep{BegelmanBlandfordRees1980}.
Many different mechanisms have been proposed over the years to evolve the binary further down to separations $\lesssim 0.1~\textrm{pc}$, such as 3-body interactions with a dense galactic stellar core and torques from a circumbinary disk, and the exact nature of such dynamic mechanisms is commonly known as the `final parsec problem' \citep{DeRosaVignali+2019}.
Once the SMBHBs reach a sub-pc separation, they inspiral by emitting GWs in the PTA frequency range and eventually merge.
Such sources have previously been predicted to be detected first as a stochastic GWB formed via the incoherent superposition of GWs coming from a population of SMBHBs, and then as individual sources that stand out above the background \citep{SesanaVecchioVolonteri2009,RosadoSesanaGair2015,MingarelliLazio+17,KelleyBlecha+2017,KelleyBlecha+2018,PolTaylor+2021,BecsyCornishKelley2022}. 
While the detection of the GWB with accurate spectral characterization could inform us about the average properties of the cosmic population of SMBHBs in our universe \citep{Taylor2021}, the detection of continuous GWs from an individual SMBHB will be a strong indicator of SMBHBs' strength of contribution to the observed GWB \citep{KelleyCharisi+2019,CharisiTaylor+2022}. 
Further, a continuous GW detection can lead to persistent multi-messenger nHz GW astronomy if we can identify its electromagnetic counterpart, which is likely to be an AGN.
Therefore, it is highly desirable to detect GWs from individual SMBHBs in the PTA data sets in the near future.

Over the years, PTA experiments have searched for and put progressively more stringent constraints on the presence of continuous GWs from individual SMBHBs in their data sets \citep{JenetLommen+2004,ZhuHobbs+2014,BabakPetiteau+2016,PereraStappers+2018,AggarwalArzoumanian+2019,FengLi+2019,NG2020_11yr_3C66B,ArzoumanianBaker+2021,FalxaBabak+2023,ArzoumanianBaker+2023,NG2023_15yr_individual,AntoniadisArumugam+2023d}.
These searches can be divided into two different categories: all-sky searches where no prior information about the source of the GWs is taken into consideration \citep[e.g.][]{BabakPetiteau+2016, AggarwalArzoumanian+2019}, and targeted searches where GWs from a particular SMBHB candidate, identified through electromagnetic observations, is searched for \citep[e.g.][]{JenetLommen+2004,NG2020_11yr_3C66B}.
For a targeted search, the sky location and the luminosity distance of the continuous GW source are fixed to their known values for the SMBHB candidate.
Furthermore, the binary period of the candidate, if obtained from electromagnetic observations, can be used to narrow down the prior for the GW frequency.
It has been shown that a targeted search can improve both the GW upper limits \citep{NG2020_11yr_3C66B} and the detection probability \citep{LiuVigeland2021} by a factor of a few as compared to an all-sky search.

Historically, most PTA searches for continuous GWs have been limited to individual SMBHBs in inspiralling quasi-circular orbits \citep[e.g.][]{AggarwalArzoumanian+2019,NG2020_11yr_3C66B}.
However, many proposed solutions for the final parsec problem rely on the binaries entering the nHz frequency regime while retaining non-negligible eccentricities \citep[e.g.][]{RyuPerna+2018}.
Furthermore, a few SMBHB candidates such as `Spikey' \citep{HuDOrazio+2020} and OJ~287 \citep{DeyGopakumar+2019} have electromagnetic signatures whose interpretation requires eccentric orbits.
Attempts at searching for eccentric SMBHBs in PTA data sets have been stymied by the prohibitive computational cost of evaluating the PTA signals induced by inspiral GWs from such sources.
This resulted in most of the past PTA searches for eccentric SMBHB sources being restricted to using a single pulsar and sub-optimal periodogram-based methods \citep{JenetLommen+2004,FengLi+2019}.
\citet{FalxaBabak+2023} and \citet{AntoniadisArumugam+2023d} searched for GWs from eccentric SMBHBs to follow up on the candidate sources identified in their circular SMBHB searches, although they did not report any detection or constraint pertaining to such sources.
The computational challenges involved in PTA searches for eccentric SMBHBs were investigated and an efficient computational framework for evaluating the PTA signals was presented in \citet{Susobhanan2023}, influenced by \citet{SusobhananGopakumar+2020}.
A pilot single-pulsar Bayesian all-sky search was also presented in \citet{Susobhanan2023} as a proof of concept.
The development of the above framework into a PTA-based eccentricity search was initially explored in \citet{Cheeseboro2021}.
In this paper, we use the framework of \citet{Susobhanan2023} to perform, for the first time, a full-PTA-scale multi-messenger targeted search for continuous GWs from an eccentric SMBHB in 3C~66B. 
We carry out this search using the NANOGrav 12.5-year (NG12.5) narrowband data set \citep{AlamArzoumanian+2020}.

3C~66B (RA: $02^{\rm h}\ 23^{\rm m}\ 11.4112^{\rm s}$, DEC: $+42^{\circ}\ 59'\ 31.385''$) is an elliptical radio galaxy with an estimated redshift of 0.02092.
This redshift corresponds to a luminosity distance ($\rm D_{\rm L}$) of 94.17 Mpc assuming a flat $\Lambda$CDM model with $H_0 = 67.66$ km/(Mpc s),  $\Omega_{m0} = 0.30966$, and $\Omega_{\Lambda0} = 0.68885$ \citep{AghanimAkrami+2020}.
The presence of an SMBHB in 3C~66B was first hypothesized by \citet{SudouIguchi+2003} using Very Long Baseline Interferometry (VLBI) measurements.
They attributed the apparent elliptical motion of 3C 66B's radio core to the orbital and precessional motion of the smaller black hole's jet. 
Based on this model, they estimated an orbital period of $1.05\pm0.03$ years and a chirp mass of $1.3\times10^{10} \Msun$.
Given the proximity of this source to our Galaxy, the estimated binary parameters implied a GW amplitude that should have been well within the detection capabilities of pulsar timing experiments.
\citet{JenetLommen+2004} used 7 years of timing data of a single pulsar, PSR B1855+09, obtained using the Arecibo telescope to search for a signal consistent with the proposed orbital period.
In the absence of a detection, they put an upper limit of $7\times10^9 \Msun$ on the chirp mass of the binary at zero eccentricity, ruling out the initially proposed model for the VLBI observations. 
\citet{IguchiOkudaSudou2010} reported flux variations of 3C~66B at 3 mm with a periodicity of 93 days, which they attributed to 
the Doppler-shifted flux modulation due to the orbital motion of the SMBHB.
Assuming the SMBHB to be in a circular orbit with an orbital period of 1.05 years as in \citet{SudouIguchi+2003}, they estimated a chirp mass of $7.9\times 10^{8} \Msun$ using this revised model, an order of magnitude less than the upper bound set by \citet{JenetLommen+2004}.
The PTA upper limits on the GWs emitted by this source were refined further by \citet{NG2020_11yr_3C66B} and \citet{ArzoumanianBaker+2023}, and the latest upper limit on the chirp mass of the 3C 66B system is $1.41\times10^9 \Msun$.
This upper limit is calculated using the NG12.5 data set \citep{AlamArzoumanian+2020} while assuming a circular binary and also accounting for the presence of a common-spectrum red noise process in the data.

While \citet{IguchiOkudaSudou2010} assumed a circular orbit to explain the periodic flux variations in 3C~66B, an eccentric SMBHB can also explain all the observed variations that suggest the presence of a binary in 3C~66B.
In fact, limits on the maximum possible eccentricity of the postulated binary in 3C~66B for a few fixed chirp masses of the system were also calculated in \citet{JenetLommen+2004} using single-pulsar data.
However, a PTA-based search for continuous GWs from an eccentric binary in 3C~66B using multiple pulsars has never been reported in the literature.
In this paper, we present a targeted search for PTA signals, induced by an eccentric SMBHB system in the radio galaxy 3C~66B, in NG12.5 data set using a Bayesian framework.
For this targeted search, we have used the orbital frequency of the proposed binary measured by \citet{SudouIguchi+2003}, the known sky location of the radio galaxy core, and the luminosity distance to the source (derived from its measured redshift) as informative priors.

This paper is arranged as follows.
In Section \ref{sec:pta-signal}, we provide a brief overview of how we model the PTA signal induced by GWs from an eccentric SMBHB, including the expressions for GW-induced pulsar timing residuals $R(t)$ (Subsection \ref{sec:pta-signal-1}), a description of the general relativistic orbital dynamics of eccentric binaries required to calculate $R(t)$ (Subsection \ref{sec:orbital-motion}), and the PTA signal parametrization for the targeted search (Subsection \ref{sec:targeted-search-params}).
In Section \ref{sec:methods}, we describe our data analysis methods, including the data set we used for the GW search (Subsection \ref{sec:NG12.5-nb}), the pulsar timing and noise models (Subsection \ref{sec:timing-model}), the prior distributions for the parameters (Subsection \ref{sec:priors}), the model comparison method (Subsection \ref{sec:model-comp}), and software pipeline used for performing the search (Subsection \ref{sec:software}).
The application of our method on a simulated data set and the corresponding results are discussed in Section~\ref{sec:simulations}.
Finally, we present the results of our targeted search for GWs from eccentric binaries in NG12.5 data set in Section \ref{sec:results} along with a summary and discussion in Section \ref{sec:discussion}.

\section{The PTA signal model for eccentric binaries}
\label{sec:pta-signal}

We begin by briefly describing the PTA signal induced by an SMBHB inspiralling along a relativistic eccentric orbit.
The computation of the PTA signal usually involves modeling the relativistic motion of the binary using a post-Newtonian (PN) formalism, where general relativistic corrections to the Newtonian binary dynamics are expressed in powers of $(v/c)^2\sim GM/(c^2r)$\footnote{A term appearing at $\mathcal{O}((v/c)^{2n})$ is usually referred to as an nPN correction.},
where $M$, $r$, and $v$ are the total mass, relative separation, and the orbital velocity of the binary, respectively, and $c$ is the speed of light in vacuum \citep{Blanchet2014}. 
The two GW polarization amplitudes $h_{+,\times}(t)$ can be expressed as functions of orbital variables, and the GW-induced pulsar timing residual (the PTA signal) $R(t)$ involves the time integrals of $h_{+,\times}(t)$.

The PTA signal originating from such a source is described in Subsection \ref{sec:pta-signal-1} and the corresponding orbital motion is described in Subsection \ref{sec:orbital-motion}, following \citet{Susobhanan2023}.
Different approaches to modeling such PTA signals may be found in \citet{JenetLommen+2004}, \citet{TaylorHuerta+2016}, and \citet{SusobhananGopakumar+2020}.

\subsection{The PTA signal}
\label{sec:pta-signal-1}

When a GW travels across our line of sight to a pulsar, it modulates the observed times of arrival (TOAs) of the pulses from that pulsar.
Such modulations are known as the PTA signal and can be expressed for a given pulsar as \citep{Detweiler1979}
\begin{equation}
R(t)=\int_{t_{0}}^{t}\left(h(t')-h(t'-\Delta_{p})\right)dt'\,,
\label{eq:pta-res}
\end{equation}
where $h$ is the GW strain (see below for precise definition), $t$ and $t'$ are coordinate times measured at the solar system barycenter (SSB), $t_0$ is an arbitrary fiducial time.
The geometric time delay $\Delta_p$ is given by
\begin{equation}
    \Delta_p = \frac{D_p}{c}(1-\cos\mu)\,,
\end{equation}
where $D_p$ is the distance to the pulsar and $\mu$ is the angle between the lines of sight to the pulsar and the GW source.

The GW strain $h(t)$ can be written as a linear combination of the two polarization amplitudes $h_{+,\times}(t)$:
\begin{equation}
     h(t) = \begin{bmatrix} F_{+} & F_{\times} \end{bmatrix}
    \begin{bmatrix}
        \cos{2 \psi} & - \sin{2 \psi} \\
        \sin{2 \psi} & \cos{2 \psi}
    \end{bmatrix}
    \begin{bmatrix} h_{+}(t) \\ h_{\times}(t) \end{bmatrix},
\end{equation}
where $\psi$ is the GW polarization angle and $F_{+,\times}$ are the antenna pattern functions that depend on the sky locations of the pulsar and the GW source (see, e.g., \citet{TaylorHuerta+2016} for explicit expressions).
Defining the functions $s_{+,\times}(t)=\int_{t_0}^t h_{+,\times}(t')dt'$, we may write $R(t)$ as
\begin{eqnarray}
R(t)=&\begin{bmatrix}F_{+} & F_{\times}\end{bmatrix}\begin{bmatrix}\cos2\psi & -\sin2\psi\\
\sin2\psi & \cos2\psi
\end{bmatrix} \nonumber\\
& \qquad \times \begin{bmatrix}s_{+}(t)-s_{+}(t-\Delta_{p})\\
s_{\times}(t)-s_{\times}(t-\Delta_{p})
\end{bmatrix}\,,
\label{eq:R(t)_px}
\end{eqnarray}
and the contributions to $R(t)$ arising from $s_{+,\times}(t)$ and $s_{+,\times}(t-\Delta_p)$ are known as the Earth term and the pulsar term, respectively.

The quadrupolar-order expressions for $s_{+,\times}(t)$, derived assuming slow advance of periapsis and orbital decay (i.e., the timescales of the advance of periapsis and the orbital decay are much longer than the orbital period), are given by \citep{JenetLommen+2004,Susobhanan2023}
\begin{subequations}   
\begin{align}
s_{+}(t)=&S_0 \varsigma\big(\left(c_{\iota}^{2}+1\right)\left(-\mathcal{P}\sin(2\omega)+\mathcal{Q}\cos(2\omega)\right) \nonumber\\
& \qquad +s_{\iota}^{2}\mathcal{R}\big)\,,\\
s_{\times}(t)=&S_0 \varsigma\,2c_{\iota}\left(\mathcal{P}\cos(2\omega)+\mathcal{Q}\sin(2\omega)\right)\,,
\end{align}
\label{eq:spx_anl}
\end{subequations}  
where
\begin{subequations}
\begin{align}
\mathcal{P} &=\frac{\sqrt{1-e_t^{2}}\,(\cos(2u)-e_t\cos(u))}{1-e_t\cos(u)}\,,\\
\mathcal{Q} &=\frac{\left(\left(e_t^{2}-2\right)\cos(u)+e_t\right)\sin(u)}{1-e_t\cos(u)}\,,\\
\mathcal{R} &= e_t \sin(u)\,.
\end{align}
\label{eq:PQ_int}
\end{subequations}
Here,
$S(t) = S_0 \varsigma(t)$ is the PTA signal amplitude at any given time $t$,
$c_\iota=\cos\iota$ and $s_\iota=\sin\iota$, where $\iota$ is the orbital inclination;
$\omega(t)$ is the argument of periastron,
and $e_t(t)$ and $u(t)$ are the time eccentricity and the eccentric anomaly, respectively.
The PTA signal amplitude is given by $S(t) = H(t)/n(t)$, where $n(t) = 2\pi/P(t)$ is the mean motion of the binary and $H(t) = GM\eta \, x(t)/ (D_{\rm L}c ^2)$ is the dimensionless GW strain amplitude;
$P(t)$ is the binary period,
$G$ is the universal gravitational constant,
$M = m_1 + m_2$ is the total mass, 
$\eta = m_1m_2/M^2$ is the symmetric mass ratio, 
$m_1$ and $m_2$ are the black hole masses, and
$D_{\rm L}$ is the luminosity distance to the binary. 
The dimensionless PN parameter $x$ is given by $x=(GM(1+k)n/c^3)^{2/3}$ where 
$k(t)$ is the advance of pericenter per orbit.
$M$ and $\eta$ relate to the chirp mass $\Mch$ as $\Mch=\eta^{3/5}M$.
Here, we have divided the time-varying amplitude of the signal into a constant amplitude $S_0 = S(t_0)$ and a time-varying part $\varsigma(t) = S(t)/S_0$ for the convenience of parametrization.
The orbital parameters appearing in equations \eqref{eq:spx_anl} and \eqref{eq:PQ_int} will be explained in greater detail in the following subsection.

\subsection{The PN-accurate orbital motion}
\label{sec:orbital-motion}

It is evident from equations \eqref{eq:R(t)_px}-\eqref{eq:PQ_int} that we need to calculate the temporal evolution of various binary orbital parameters in order to determine $R(t)$.
The orbital evolution of a relativistic binary can be split into two parts: the conservative part that takes into account the advance of periapsis of the orbit, and the reactive evolution due to emission of GWs from the system.
We use the PN-accurate quasi-Keplerian parametrization to describe the conservative dynamics of our relativistic eccentric binary systems \citep{DamourDeruelle1985,MemmesheimerGopakumarSchafer2004}.
We begin our description by defining the mean anomaly $l$ as
\begin{equation}
    l(t) = \int_{t_0}^t n(t')\, dt'\,,
\end{equation}
and expressing the eccentric anomaly $u$ implicitly as a function of $l$ via the PN-accurate Kepler equation \citep{MemmesheimerGopakumarSchafer2004,BoetzelSusobhanan+2017}
\begin{equation}
l=u-e_{t}\sin u+\mathfrak{F_{t}}(u)\,,
\label{eq:kepler}
\end{equation}
where $\mathfrak{F}_t(u)$ is a periodic function of $u$ that first appears at the 2PN order.
The angular coordinate in the orbital plane (`orbital phase') $\phi$ can be written as
\begin{equation}
\phi=\varpi+l+(1+k)(f-l)+\mathfrak{F}_{\phi}(u)\,,
\label{eq:phi}
\end{equation}
where $f$ is the true anomaly given by
\begin{equation}
    f = 2\arctan\left[\sqrt{\frac{1+e_\phi}{1-e_\phi}}\tan\frac{u}{2}\right]\,,
\label{eq:true_anomaly}
\end{equation}
$k$ is the advance of periastron per orbit, $\varpi$ is the periastron angle defined by
\begin{equation}
    \varpi = \int_{t_0}^t k(t') n(t')\,dt'\,,
\label{eq:gamma}
\end{equation}
$e_\phi$ is known as the angular eccentricity, and $\mathfrak{F}_\phi(u)$ is a periodic function of $u$ first appearing at the 2PN order.
The argument of periastron $\omega=\phi-f$, and is not to be confused with $\varpi$.\footnote{The periastron angle is denoted by $\gamma$ in \citet{SusobhananGopakumar+2020} and \citet{Susobhanan2023}. We denote this quantity by $\varpi$ in this paper to avoid confusion with the red noise power law indices appearing in Section \ref{sec:timing-model}.}
We use the 3PN-accurate expressions for $k$, $e_\phi$, $\mathfrak{F}_t$, and $\mathfrak{F}_\phi$ involving $x$, $e_t$, and $\eta$ found in  \citet[supplementary material]{BoetzelSusobhanan+2017}.

In the absence of galactic environmental influences, the orbital period and eccentricity of the binary decrease with time due to the emission of GWs from the system, and this reactive evolution of the orbit can be incorporated into our framework by allowing $n$ and $e_t$ to vary slowly as functions of time.
Accurate up to the leading quadrupolar (2.5PN) order, the orbital evolution can be expressed as a system of four coupled differential equations \citep{DamourGopakumarIyer2004}
\begin{subequations}
\begin{align}
\frac{dn}{dt} & =\frac{1}{5}\left(\frac{GMn}{c^{3}}\right)^{5/3}\eta n^{2}\frac{\left(96+292e_{t}^{2}+37e_{t}^{4}\right)}{\left(1-e_{t}^{2}\right)^{7/2}}\,,\\
\frac{de_{t}}{dt} & =\frac{-1}{15}\left(\frac{GMn}{c^{3}}\right)^{5/3}\eta ne_{t}\frac{\left(304+121e_{t}^{2}\right)}{\left(1-e_{t}^{2}\right)^{5/2}}\,,\\
\frac{d\varpi}{dt} & =k\,n\,,\\
\frac{dl}{dt} & =n\,.
\end{align}
\label{eq:orbit-odes}
\end{subequations}
In this work, we solve the above system of differential equations using the analytic solution provided in \citet{SusobhananGopakumar+2020}  to get the temporal evolution of $n$, $e_t$, $\varpi$, and $l$, starting with certain initial conditions $(n_0, e_0, \varpi_0, l_0)$ defined at $t_0$.
Thereafter, we use equations \eqref{eq:kepler}, \eqref{eq:phi}, and \eqref{eq:true_anomaly} to calculate the evolution of $u$ and $\omega$ as a function of time.
Now we have all the inputs to calculate the PTA signal $R(t)$ using equations \eqref{eq:R(t)_px}, \eqref{eq:spx_anl}, and \eqref{eq:PQ_int}.

\subsection{Parametrizing the PTA signal model for a targeted search}
\label{sec:targeted-search-params}

To calculate the PTA signal due to GWs from an individual SMBHB for a particular pulsar, we need information about various quantities related to the GW source and the pulsar.
These include the sky position of both the GW source (RA\textsubscript{gw}, DEC\textsubscript{gw}) and the pulsar (RA\textsubscript{psr}, DEC\textsubscript{psr}), required to calculate the antenna pattern functions $F_{+,\times}$.
The other parameters for the eccentric SMBHB GW model, gleaned from subsections \ref{sec:pta-signal-1} and \ref{sec:orbital-motion}, are
the PTA signal amplitude ($S_0$) at a reference time $t_0$,
the projection angles ($\psi$, $\iota$),
the binary mass parameters ($M$, $\eta$), and
the initial condition for the orbital evolution ($n_0$, $e_{0}$, $\varpi_0$, $l_0$) given at $t_0$.
{Further, the pulsar distance $D_p$ also enters the model if the pulsar term contributions are included, where $p$ is an index denoting the pulsar.}

In the case of a targeted search, the GW source coordinates and the redshift ($z$) are usually well-known from electromagnetic observations, and the luminosity distance $D_{\rm L}$ to the source can be computed from $z$ by assuming a cosmological model.
For some SMBHB candidates (e.g., 3C~66B, OJ~287; \citealt{DeyValtonen+2018}), the orbital period $P_0$ (and therefore $n_0$) at some $t_0$ is also accurately measured from the electromagnetic data.
Additionally, given $S_0$, $n_0$, $e_0$, $\eta$, and $D_{\rm L}$, one can compute $M$ by numerically solving the equation\footnote{We apply the Newton-Raphson method. This requires the derivative $\frac{\partial S_0}{\partial M}\approx \frac{(19 k_0+15) S_0}{9 (k_0+1) M}$, where $k_0$ denotes the value of $k$ at $t_0$.}
\begin{equation}
    S_0 = \frac{GM\eta}{n_0 D_{\rm L} c^2}\left(\frac{GM(1+k(M,\eta,n_0,e_0))n_0}{c^3}\right)^{2/3}\,.
    \label{eq:S0_M_relation}
\end{equation}
We can also rewrite this equation as
\begin{equation}
    S_0 = \frac{(G\Mch)^{5/3}}{D_{\rm L} c^4\, n_0^{1/3}} (1+k(M,\eta,n_0,e_0))^{2/3}\,,
    \label{eq:S0_Mch_relation}
\end{equation}
where $\Mch = M \eta^{3/5}$ is the chirp mass of the binary.
Therefore, the unknown or free independent model parameters related to the GW source for a targeted search include $(S_0,\psi,\iota,\eta,e_0,\varpi_0,l_0)$.

Regarding the pulsar-related parameters, the pulsar sky positions are very accurately known from pulsar timing.
The pulsar distances are usually not well-constrained by electromagnetic observations and hence we treat the pulsar distance for each pulsar as an unknown parameter.
In addition, the initial orbital phase parameters $l_p$ and $\varpi_p$ for the pulsar term are treated as independent from $l_0$ and $\varpi_0$ due to the poor constraint on the pulsar distances and the errors arising from restricting the reactive orbital evolution to the leading PN order.
To summarize, we use, along with the 7 GW-source parameters mentioned above, $(D_p,\varpi_p,l_p)$ for each pulsar as free model parameters for our targeted search.
Obviously, the pulsar-specific parameters are omitted in the case of an Earth term-only search, where we consider only the Earth term contributions to the PTA signal and ignore the pulsar term contributions.

\section{Data analysis methods}
\label{sec:methods}

In this section, we begin by describing briefly the pulsar timing data set used in this paper, along with the full model 
we used to describe the pulsar timing residuals.
The priors we have used for different free model parameters are also discussed in detail.
The software pipeline used in our analysis is also discussed.

\subsection{The NANOGrav 12.5-year narrow-band data set}
\label{sec:NG12.5-nb}

As mentioned earlier, we have used the NANOGrav 12.5-year (NG12.5) narrowband data set to perform our targeted search for continuous GWs from an eccentric SMBHB in 3C~66B. 
This data set consists of TOAs and timing models for 47 pulsars obtained using the Green Bank Telescope and the Arecibo Telescope from 2004 to 2017 \citep{AlamArzoumanian+2020}.
We excluded from our analysis PSRs J1946+3417 and J2322+2057 due to their short observation span (less than 3 years), and PSR J1713+0747 due to the presence of two chromatic timing events within the data span that were identified to be of non-GW (possibly pulsar magnetospheric or interstellar medium) origin \citep{LamEllis+2018}.
The earliest and latest TOAs in this data set are MJDs 53216 (2004-30-07) and 57933 (2017-06-29) respectively, making the data span $\sim 12.92$ years. 
Each TOA is transformed to the solar system barycenter (SSB) frame using the DE438 solar system ephemeris\footnote{\url{https://naif.jpl.nasa.gov/pub/naif/JUNO/kernels/spk/de438s.bsp.lbl}}.
The dispersion measure (DM) variations present in the TOAs are corrected by applying DMX parameters, which provide a piecewise constant model thereof. 

\subsection{Model for the timing residuals}
\label{sec:timing-model}

Pulsar timing residuals represent the deviations in the observed TOAs from the ones predicted by a timing model. 
The timing residuals $r$ in the vicinity of the maximum-likelihood values of the timing model parameters can be expressed as a sum of different components as follows:
\begin{equation}
    r = \boldsymbol{\mathfrak{M}}\boldsymbol{\epsilon} + n_{\text{WN}} + n_{\text{IRN}} + n_\CURN + R\,,
    \label{eq:full_signal_model}
\end{equation}
where $\boldsymbol{\mathfrak{M}}$ is the pulsar timing design matrix and $\boldsymbol{\epsilon}$ is a vector that represents the deviations in timing model parameters from their maximum-likelihood values arising due to the presence of signal components that were unaccounted for in the initial timing model.
The linearized timing model deviations $\boldsymbol{\epsilon}$ are analytically marginalized to reduce the number of dimensions of the parameter space \citep{vanHaasterenLevin+2009}.
The term $n_\text{WN}$ represents the white noise (WN) that can arise due to pulse jitter, instrumental noise, radio frequency interference, polarization miscalibration, etc \citep{NG2023_15yr_noise}.
The individual pulsar red noise (IRN) term denoted by $n_\text{IRN}$ models the deviations in the TOAs due to rotational irregularities of the pulsar.
The term $n_\CURN$ represents a common uncorrelated red noise (CURN) while $R$ represents the PTA residuals induced by GWs from individual eccentric SMBHB described in Section \ref{sec:pta-signal}.

The WN term $n_{\text{WN}}$ is modeled using three phenomenological parameters EFAC, EQUAD, and ECORR, each for every telescope receiver/backend combination.
The EFAC parameters scale the TOA measurement uncertainties by a multiplicative factor, the EQUAD parameters add to the measurement uncertainties in quadrature, and the ECORR parameters model the correlated white noise between narrowband TOAs derived from the same observation \citep{NG2023_15yr_noise}. 

The IRN for each pulsar is modeled as a Gaussian red noise process with a power-law spectral density
\begin{equation}
    P_{\text{IRN},p}(f) = \frac{A^2_{p}}{12 \pi^2} \left( \frac{f}{f_{\text{ yr}}}\right)^{-\gamma_{p}}\, {\text{yr}^3} ,
\label{eq:IRN_pl}
\end{equation}
with amplitude $A_{p}$ and spectral index $\gamma_{p}$, where $p$ is an index denoting the pulsar and $f_\text{yr}=1~\text{yr}^{-1}$.
We model the IRN of a pulsar as a Fourier sum consisting of 30 linearly spaced frequency bins ranging from $1/T_{\text{span}}$ to $30/T_{\text{span}}$, where $T_{\text{span}}$ represents the total observation time span of the data set.

In addition to the IRN, \citet{NG2020_12yr_GWB} showed the presence of a common-spectrum red noise process $n_\CURN$ in the NG12.5 data set, albeit without any conclusive detection of spatial correlations.
Recently, this common process was shown to be an early hint of the GWB, which manifests as an HD-correlated common-spectrum process in the NANOGrav 15-year data set \citep{NG2023_15yr_GWB}.
In this work, we treat $n_\CURN$ as a spatially uncorrelated common-spectrum process since spatial correlations are not detected in the NG12.5 data set.
We model $n_\CURN$ as a Gaussian process with a power-law spectrum as
\begin{equation}
    P_{\CURN}(f) = \frac{A^2_{\CURN}}{12 \pi^2} \left( \frac{f}{f_{\rm yr}}\right)^{-\gamma_{\CURN}}\, {\rm yr^3}\,,
\label{eq:CURN_pl}
\end{equation}
where $A_{\CURN}$ is the amplitude and $\gamma_{\CURN}$ is the spectral index.
We model the CURN as a Fourier sum consisting of 5 linearly spaced frequency bins ranging from $1/T_{\text{span}}$ to $5/T_\text{span}$ to be consistent with the NG12.5 GWB analysis \citep{NG2020_12yr_GWB}.
Overall, the models we use for the WN, IRN, and the CURN are same as those from the NG12.5 GWB analysis.

\subsection{Priors}
\label{sec:priors}

For IRN and the CURN parameters, we use priors that are identical to those used in the NG12.5 GWB analysis \citep{NG2020_12yr_GWB}. 
This corresponds to log uniform prior for the amplitudes and uniform prior for the spectral indices, namely $\log_{10}A \in \text{U}[-18, -11]$ and $\gamma \in \text{U}[0, 7]$, where $A$ and $\gamma$ represent the amplitudes and spectral indices appearing in equations \eqref{eq:IRN_pl} and \eqref{eq:CURN_pl}.
Further, we fix the white noise parameters at their maximum likelihood values obtained from single-pulsar analysis as is standard practice in PTA analyses \citep[e.g.][]{ArzoumanianBaker+2023}.
This amounts to 90 free noise parameters in total: two IRN parameters per pulsar and two CURN parameters.

\begin{table*}[t]
    \centering
    \begin{tabular}{c|c|c|c}
     \hline
\textbf{Parameter} & \textbf{Description} & \textbf{Unit}  & \textbf{Prior}         \\ \hline
RA\textsubscript{gw} & Right ascension of the GW source & rad & 	Constant[0.62478]\\
DEC\textsubscript{gw} & Declination of the GW source &rad & 	Constant[0.681897]\\
$t_0$  & Fiducial time & MJD               & Constant[52640]\\
$P_0$  & Orbital period at fiducial time & yr                & Constant[1.05]\\
$D_{\rm L}$  & Luminosity distance to the GW source & Mpc               & Constant[94.17]\\
\hline
$\log_{10}S_0$  & PTA signal log-amplitude at fiducial time & s               & Uniform[-12, -6] $\times ~ V(\log_{10} S_0, e_0, \eta)$ \\
 & & & LinearExp[-12, -6] $\times ~ V(\log_{10} S_0, e_0, \eta)$ \\ 
$\psi$        & GW Polarization angle & rad             & Uniform[0, $\pi$]  \\ 
$\cos\iota$             & Cosine of orbital inclination &              & Uniform[-1, 1] \\ 
$\eta$             & Symmetric mass ratio &                 & Uniform[0.001, 0.25]  $\times ~ V(\log_{10} S_0, e_0, \eta)$ \\ 
$e_0$               & Eccentricity at fiducial time &                 & Uniform[0.001, 0.8] $\times ~ V(\log_{10} S_0, e_0, \eta)$\\ 
$\varpi_0$             & Periastron angle at fiducial time (for Earth term) & rad             & Uniform[0, $\pi$] \\ 
$l_0$             & Mean anomaly at fiducial time (for Earth term) & rad             & Uniform[0, $2\pi$] \\ \hline
$D_p$       & Distance to the pulsar & kpc              & PsrDistPrior  \\ 
$\varpi_p$       & Periastron angle at fiducial time (for Pulsar term)  & rad              & Uniform[0, $\pi$]  \\ 
$l_p$       & Mean anomaly at fiducial time (for Pulsar term)  & rad              & Uniform[0, $2\pi$]  \\ \hline
    \end{tabular}
    \caption{Prior distributions for the PTA signal parameters. 
    The source coordinates of our GW source 3C~66B are taken from NASA/IPAC Extragalactic Database (\url{http://ned.ipac.caltech.edu/}).
    $P_0$ is taken from \citet{SudouIguchi+2003} and $D_{\rm L}$ is estimated from the redshift assuming a flat Universe with cosmological parameters given in \citet{AghanimAkrami+2020}.
    The fiducial time $t_0$ is an approximate reference time for the end of \citet{SudouIguchi+2003} observations.
    `PsrDistPrior' represents Normal distributions truncated at 0 obtained using parallax and DM-based distance estimates (see text for details).
    In addition to the above single-parameter priors, the joint prior distribution of $S_0$, $\eta$, and $e_0$ is modified by certain validation criteria $V(\log_{10} S_0, e_0, \eta)$ as described in the text.}
    \label{tab:priors}
\end{table*}

The prior distributions for the eccentric SMBHB parameters used in our analysis are listed in Table \ref{tab:priors} along with the fixed parameters known from the binary model of 3C~66B \citep{SudouIguchi+2003, IguchiOkudaSudou2010}.
{We fix the sky position of the GW source to the known optical location of 3C~66B. The uncertainty in the sky position of 3C~66B should not significantly affect our results as it is very small compared to the spatial resolution sensitivity of our data. 
This can be inferred from previous all-sky searches for continuous GWs \citep[e.g.,][]{ArzoumanianBaker+2023,NG2023_15yr_individual} where it is seen that the upper limits do not change rapidly with sky position.
The orbital period $P_0$ of the binary in 3C~66B estimated from electromagnetic observations is $1.05 \pm 0.03$ years.
We treat $P_0$ as a constant in our analysis because the above-mentioned uncertainty in the period corresponds to a GW frequency uncertainty of $\sim 0.9$ nHz, which is smaller than the frequency resolution of $\sim 2.5$ nHz for the NG12.5 data set.}
The luminosity distance $D_{\rm L}$ of the GW source is estimated using the redshift ($z=0.02092$) assuming a flat Universe with cosmological parameters given in \citet{AghanimAkrami+2020}, available as \texttt{Planck18} in \texttt{astropy} \citep{Price-WhelanSipocz+2018}.

We use two types of priors for the PTA signal amplitude $S_0$ following the standard practice adopted in PTA searches: the detection analysis is performed using a uniform prior on $\log_{10}S_0$ and the upper limit analysis is performed using a uniform prior on $S_0$ (denoted as `LinearExp' for $\log_{10}S_0$ in Table \ref{tab:priors}).
Although $\eta\in(0,0.25]$ and $e_0\in[0,1)$ by definition, we set their lower bounds to small non-zero numbers to avoid division by zero errors.
We also restrict $e_t<0.8$ because, for large values of $e_t$, the angular eccentricity $e_\phi$ may go above 1 for highly relativistic systems, and this renders the quasi-Keplerian parametrization invalid.
This is discussed further in Section~\ref{sec:results}.
The upper bound of $\psi$ is set to $\pi$ as it enters the PTA signal as $\sin2\psi$ and $\cos2\psi$, and similar priors apply for $\varpi_0$ and $\varpi_p$ also.

To set the priors for $D_p$ for each pulsar, we use the pulsar distance obtained from the parallax measurement if available.
Otherwise, we use the NE2001 galactic DM model \citep{CordesLazio2002} to estimate the distance from the DM of the pulsar (see Section 2.3.4 of \citet{ArzoumanianBaker+2023} for a detailed discussion).
The pulsar distances and uncertainties used in our analysis can be found in Table~2 of \citet{ArzoumanianBaker+2023}.
We use a truncated Normal distribution to ensure that $D_p>0$, and scale the uncertainties of DM-based $D_p$ estimates by a factor of 2 to account for systematic errors inherent in such measurements. 
This is denoted as `PsrDistPrior' in Table \ref{tab:priors}.
The pulsar sky locations are accurately known from pulsar timing. 
We treat them as fixed parameters while computing the PTA signals since their uncertainties are too small to affect our results.

\begin{figure*}
    \centering
    \includegraphics[scale=0.63]{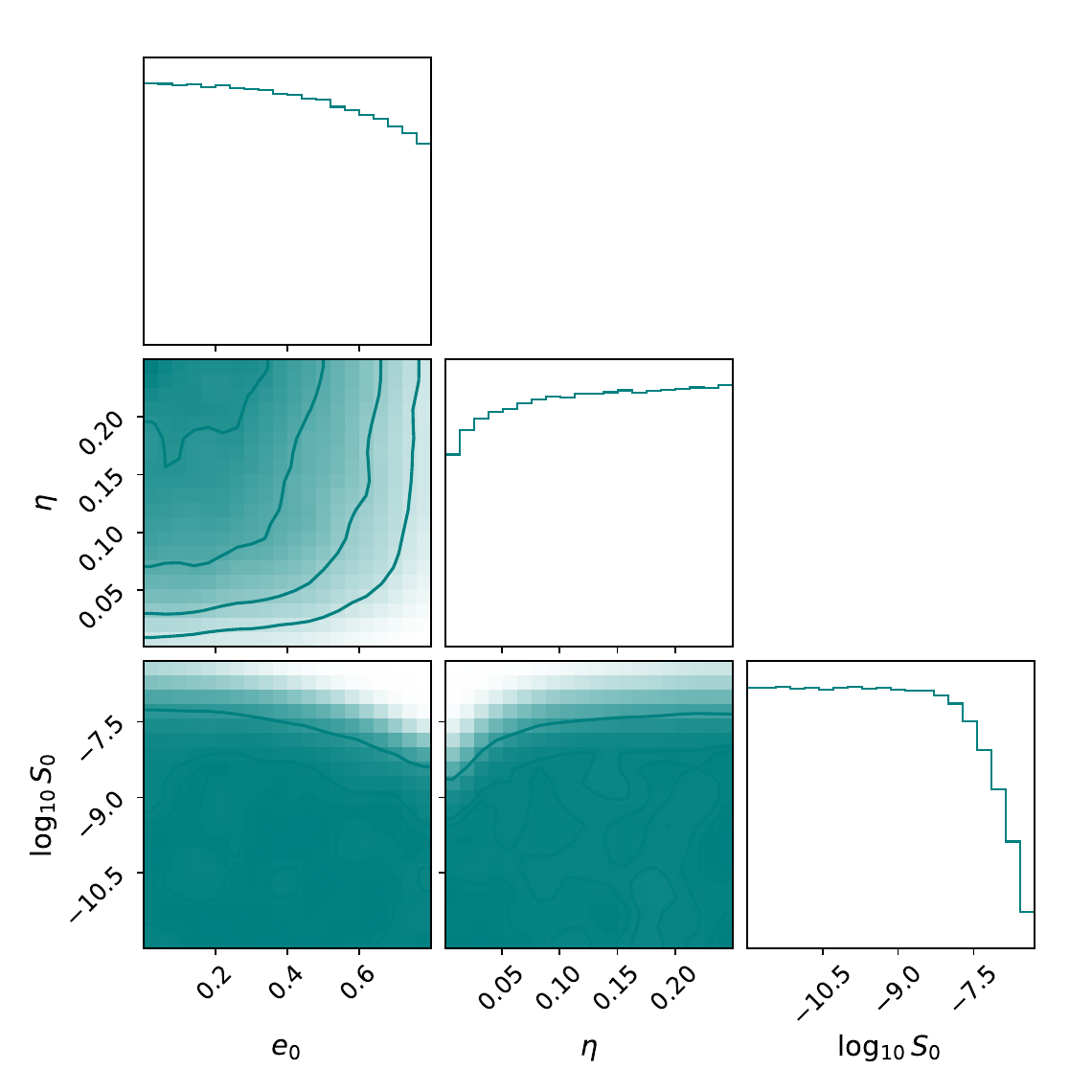}
    \caption{Corner plot of the joint prior distribution of $S_0$, $\eta$, and $e_0$ after the validation criteria $V(\log_{10} S_0, e_0, \eta)$ have been applied.
    The samples are drawn from the prior using rejection sampling.
    }
    \label{fig:validated-prior}
\end{figure*}

In addition to the above single-parameter priors, we set the prior to zero for the combinations of parameters that fail either of the following criteria. 
\begin{enumerate}
    \item The binary should not merge within the data span (see \citet{SusobhananGopakumar+2020} for an expression for a quadrupolar-order approximation for the merger time).
    \item The post-Keplerian parameters $k<0.5$ and $e_\phi<1$ throughout the evolution of the binary within the data span.
\end{enumerate}
As the binary approaches the merger, the quadrupolar approximation becomes inadequate and a more sophisticated inspiral-merger-ringdown description of the PTA signal becomes necessary.
Since we lack such a description currently, we must exclude this regime from our analysis.
Furthermore, the quasi-Keplerian description breaks down as the binary becomes more and more relativistic, and we exclude this regime from our analysis by employing the somewhat arbitrary condition $k<0.5$.
For a given $P_0$ and $D_{\rm L}$ of the SMBHB, the evolution of the binary is uniquely determined by $S_0$, $e_0$, and $\eta$ as the total mass $M$ can be calculated using equation~\eqref{eq:S0_M_relation}.
Therefore, we define a validation function $V(\log_{10} S_0, e_0, \eta)$ such that it is set to 1 if the combination of $S_0$, $e_0$, and $\eta$ values satisfies the criteria mentioned above, otherwise 0.
We multiply this validation function with the otherwise uniform prior distributions for $\log_{10} S_0$, $e_0$, and $\eta$, as shown in Table \ref{tab:priors}, to make sure the prior is zero in case our description of the PTA signal is inadequate.
The effect of these extra criteria is that they alter the joint prior distribution of $S_0$, $\eta$, and $e_0$, and this is shown as a corner plot in Figure \ref{fig:validated-prior}.

It is evident from Figure~\ref{fig:validated-prior} that the upper limit of the valid prior space for $\log_{10} S_0$ depends on $e_0$ and $\eta$, and it is lower for large $e_0$ and small $\eta$ values.
This is expected because for the same PTA signal amplitude $S_0$ (which is proportional to $\Mch^{5/3}$ at the leading order; see equation~\eqref{eq:S0_Mch_relation}), a  system with higher eccentricity is more relativistic due to the lower relative separation at the periapsis.
Further, for the same PTA signal amplitude $S_0$, a binary with a very small value of $\eta$ demands a very large value of the total mass (as evident from equation~\eqref{eq:S0_M_relation}), which in turn results in a large value of the rate of advance of periapsis $k$.
These properties of the valid joint prior distributions can affect our analysis, especially the calculation of the upper limit of $S_0$ in case of non-detection, in certain regions of the parameter space.
This will be further discussed in Section~\ref{sec:results}.
We stress that these criteria represent the limitations of our PTA signal model rather than being physical.
We also note in passing that a similar constraint on the prior distribution based on the innermost stable circular orbit frequency was employed in the context of circular binary searches in \citet{AggarwalArzoumanian+2019} and \citet{ArzoumanianBaker+2023}.

\subsection{Model comparison}
\label{sec:model-comp}
In order to determine the significance of the detection of a PTA signal due to an eccentric binary, we compare the model $\mathcal{H}_1$ = Timing Model + WN + IRN + CURN + eccentric PTA signal against the model $\mathcal{H}_0$ = Timing Model + WN + IRN + CURN by computing the Bayes factor using the Savage-Dickey formula \citep{Dickey1971}.
We are able to apply the Savage-Dickey formula here because $\mathcal{H}_1$ and $\mathcal{H}_0$ are nested models, with $\mathcal{H}_0$ being the special case of $\mathcal{H}_1$ when $S_0=0$.
The Bayes factor is given in this case by
\begin{align}
    \mathcal{B}_{10} = \frac{p[S_0=0|\mathcal{H}_1]}{p[S_0=0|\mathcal{D},\mathcal{H}_1]}\,,
\end{align}
where $\mathcal{D}$ represents the data, and $p[S_0=0|\mathcal{H}_1]$ and $p[S_0=0|\mathcal{D},\mathcal{H}_1]$ represent the values of the marginalized prior distribution and the posterior distribution at $S_0=0$, respectively.
The uncertainty in this estimate is given by $\mathcal{B}_{10}/\sqrt{N_0}$ where $N_0$ is the number of samples in the lowest amplitude bin.

\subsection{Search pipeline implementation}
\label{sec:software}

We use the \gwecc\footnote{\url{https://github.com/abhisrkckl/GWecc.jl}} package \citep{Susobhanan2023} to evaluate the PTA signal and the \texttt{ENTERPRISE}\footnote{\url{https://github.com/nanograv/enterprise}} package \citep{EllisVallisneri+2017} to evaluate the single-parameter prior distributions and the likelihood function.
In \gwecc, the \texttt{eccentric\_pta\_signal\_target} function computes the PTA signal parametrized for a targeted search, the \texttt{gwecc\_target\_block} function provides a corresponding deterministic signal object that can be used with \texttt{ENTERPRISE}, the \texttt{PsrDistPrior} class implements the informative pulsar distance priors, and \texttt{validate\_params\_target} function provides the criteria for defining the modified joint prior distribution as described in Subsection \ref{sec:priors}. 
The Hermite spline-based method described in \citet{Susobhanan2023} is used to accelerate the PTA signal computation.

We use \texttt{PTMCMCSampler}\footnote{\url{https://github.com/jellis18/PTMCMCSampler}} \citep{EllisvanHaasteren2017} to draw samples from the posterior distribution.
\texttt{PTMCMCSampler} implements adaptive Metropolis (AM), single component adaptive Metropolis (SCAM), and differential evolution (DE) proposal distributions in combination with parallel tempering to effectively draw samples from a given posterior distribution.
\texttt{PTMCMCSampler} also supports user-defined proposal distributions along with tunable weights for each proposal distribution.
We use the following proposals available in the \texttt{enterprise\_extensions}\footnote{\url{https://github.com/nanograv/enterprise\_extensions}} package \citep{TaylorBaker+2021}: draws from the single-parameter priors (\texttt{JumpProposal.draw\_from\_prior} and \texttt{JumpProposal.draw\_from\_par\_prior}), and draws from empirical distributions (\texttt{JumpProposal.draw\_from\_empi
rical\_distr}) for IRN and CURN parameters. 
In addition, we use parallel tempering with four geometrically spaced temperatures.
\texttt{ENTERPRISE}, \texttt{enterprise\_extensions}, and \texttt{PTMCMCSampler} are described in \citet{JohnsonMeyers+2023}.
The search and visualization scripts used in this work can be found at \url{https://github.com/lanky441/NG12p5_3C66B_GWecc}.

\section{Simulations}
\label{sec:simulations}

Before applying our methods to the NG12.5 data set, we perform targeted searches for eccentric PTA signals in simulated data sets to test our methods' performance and reliability.
Inspired by \citet{CharisiTaylor+2023}, we carry out two separate analyses: an \textit{Earth term-only} (E-only) analysis (pulsar term is excluded from the searched-for signal model) and an \textit{Earth+Pulsar terms} (E+P) analysis (pulsar term is included in the searched-for signal model).

\begin{figure*}
    \centering
    \includegraphics[width=1\textwidth]{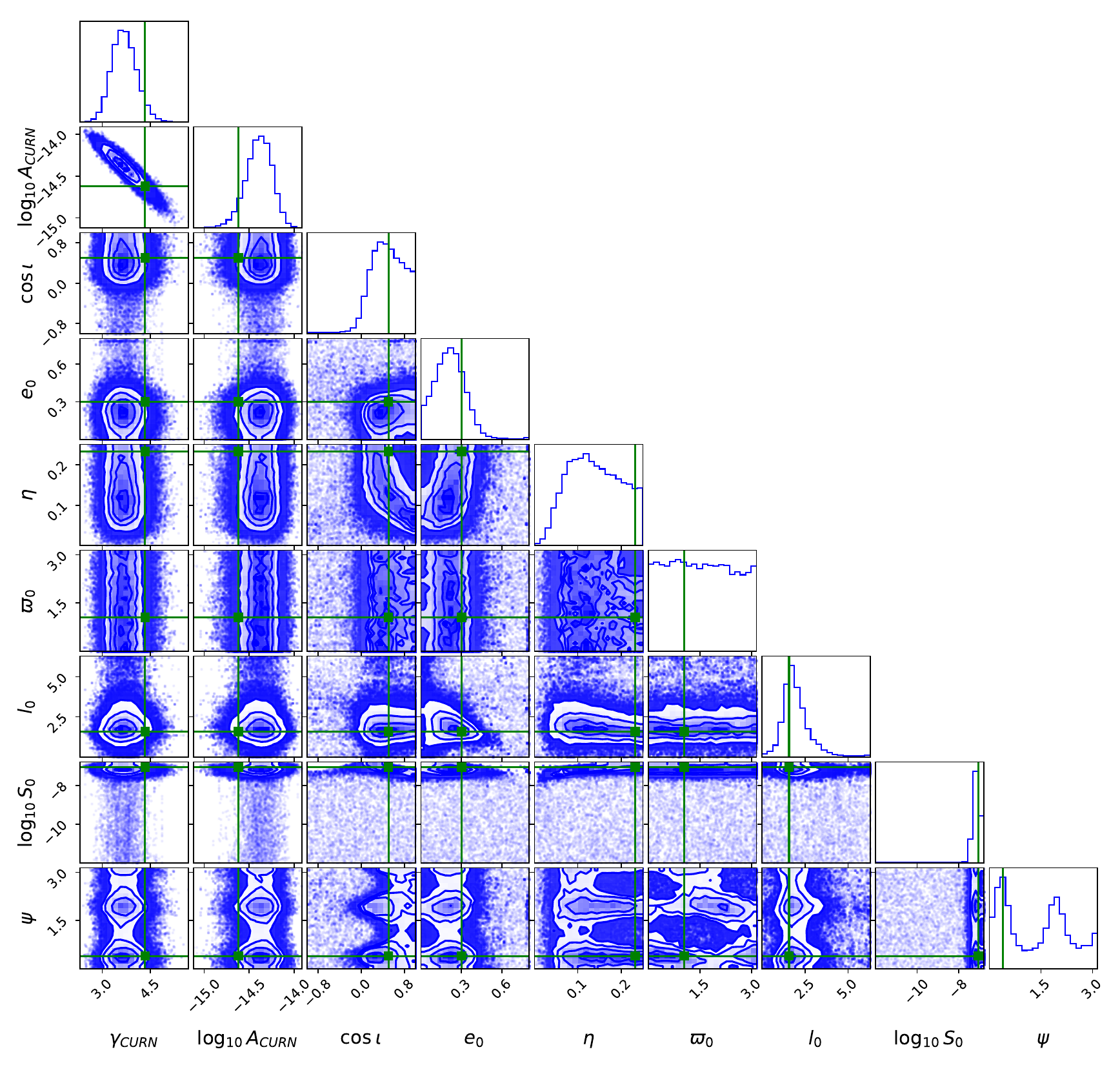}
    \caption{Posterior distributions for the CURN parameters ($\gamma_{\CURN}$ and $\log_{10}A_{\CURN}$) and the eccentric SMBHB parameters, marginalized over other parameters, for our Earth term-only (E-only) search for GWs from eccentric binary in a simulated data set.
    {The sky location and period of the simulated GW source are taken from Table \ref{tab:priors}, the binary masses are taken from \citet{IguchiOkudaSudou2010}, and the luminosity distance is taken to be one-fifth the value given in Table \ref{tab:priors}.}
    The green lines indicate the true values of the parameters for the injected signals used to create the simulated data set.
    }
    \label{fig:sim_eonly}
\end{figure*}

\begin{figure*}
    \centering
    \includegraphics[width=1\textwidth]{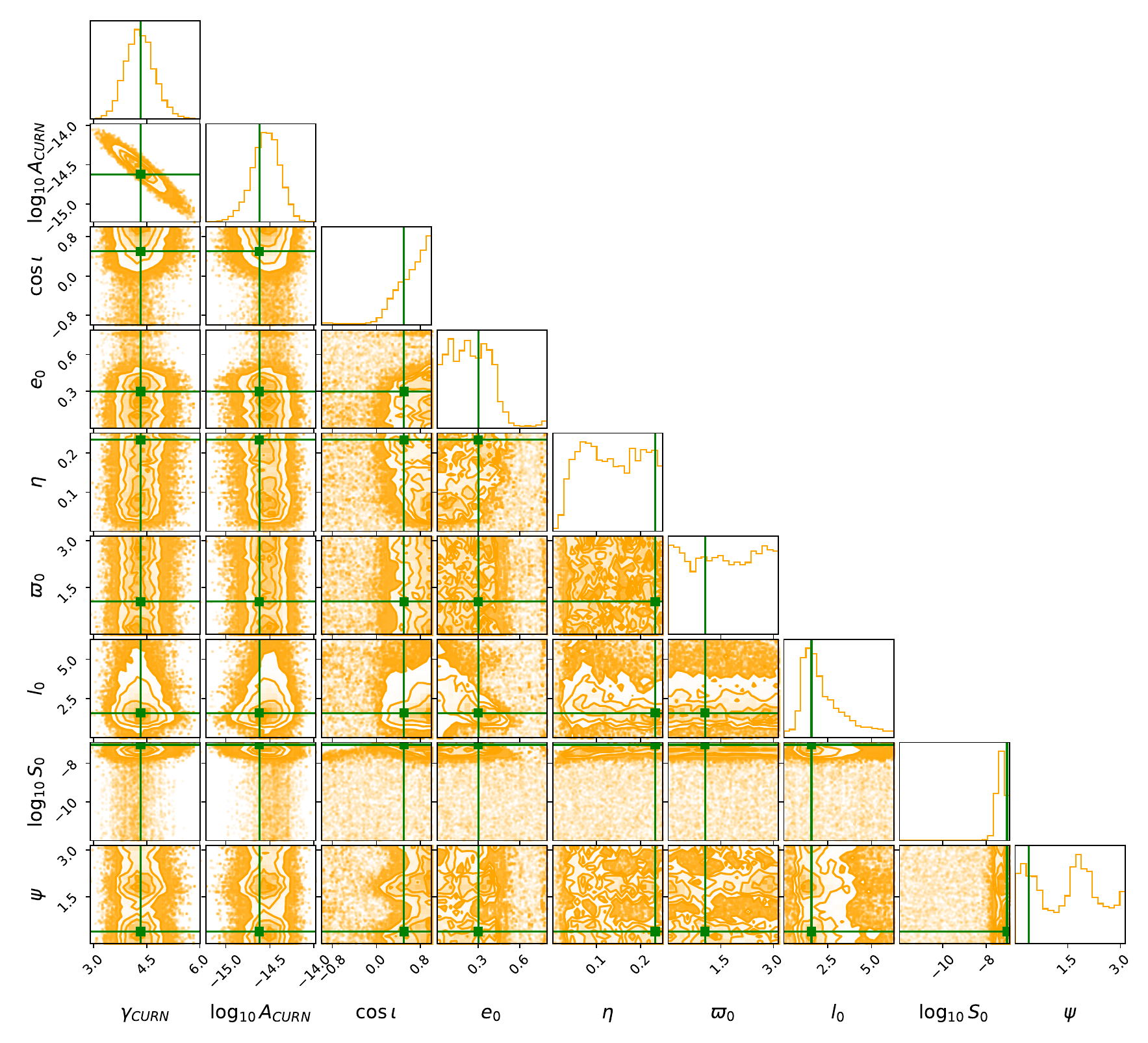}
    \caption{Posterior distributions of the parameters for our Earth+Pulsar term (E+P) search in simulated data.
    {See the caption of Figure~\ref{fig:sim_eonly} for details on the underlying simulated dataset and plotting conventions.}}
    \label{fig:sim_ep}
\end{figure*}

\begin{figure*}
    \centering
    \includegraphics[width=0.6\textwidth]{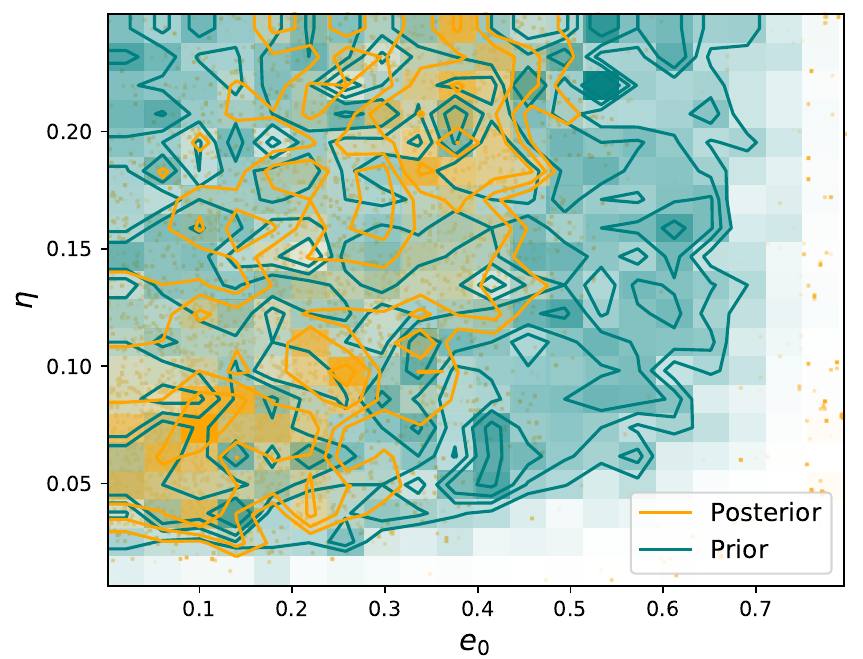}
    \caption{Comparison between the valid joint prior distribution and the 2D marginalized posterior distribution of $e_0$ and $\eta$ for E+P search of GWs in simulated data.
    {The underlying simulated data is the same as in Figure \ref{fig:sim_eonly}.}
    We see that the upper limit of the posterior distribution for $e_0$ is not limited by the limitations in the prior.
    However, the lower limit of the posterior distribution of $\eta$ may be limited by the prior.
    }
    \label{fig:sim_e+p-prior-comp}
\end{figure*}

The simulated data sets used in our study correspond to the ten pulsars with the highest dropout factors \citep{AggarwalArzoumanian+2019} as estimated in the NG12.5 GWB search \citep{NG2020_12yr_GWB}, and they are generated using the \texttt{libstempo} package\footnote{\url{https://github.com/vallis/libstempo}} \citep{Vallisneri2020}, a python wrapper for \texttt{tempo2}\footnote{\url{https://bitbucket.org/psrsoft/tempo2}} \citep{HobbsEdwardsManchester2006}.
We begin by creating an ideal noise-free realization of the pulsar TOAs by subtracting the timing residuals from the measured TOAs taken from the NG12.5 narrow-band dataset. 
We inject a WN realization with the same measurement uncertainties as in the NG12.5 narrow-band dataset, with EFAC=1 and EQUAD=0. 
We also inject a CURN realization with $A_\text{CURN} = 2.4 \times 10^{-15}$ (the GWB amplitude estimated from the NANOGrav 15-year data set \citep{NG2023_15yr_GWB}), and $\gamma_\text{CURN} = 4.33$ as expected of a GWB from a population of circular SMBHBs undergoing GW emission-induced orbital evolution \citep{Phinney2001}.
We do not include IRN realizations in the simulated data for the sake of simplicity and to ensure that the analysis finishes within a reasonable time.
{Thereafter, we inject an eccentric PTA signal from a 3C~66B-like SMBHB, i,e., an SMBHB at the same sky location of 3C~66B with the same proposed binary parameters but a different distance.
Both the Earth and the Pulsar terms contributions are included while injecting the PTA signal.}
Finally, we fit the original timing model to the signal-injected TOAs in a maximum-likelihood way and save the resulting post-fit timing model and TOAs as \texttt{par} and \texttt{tim} files.
Note that we use the same simulated data set for both E-only and E+P analyses since a physical PTA signal will include both the Earth term and the Pulsar term contributions.

The aforementioned injected eccentric SMBHB signal is computed using the binary period and sky location as listed in Table~\ref{tab:priors}, {and the binary mass estimates given in \citet{IguchiOkudaSudou2010}}.
{In order to explore the performance and reliability of our pipeline for different signal strengths, we created simulations with different GW source luminosity distance values, namely $D_\text{L}^\text{true}$, $D_\text{L}^\text{true}/5$, and $D_\text{L}^\text{true}/10$, where $D_\text{L}^\text{true}$ is the luminosity distance value listed in Table \ref{tab:priors}. 
}


We performed the E-only and E+P searches in the simulated data sets using our pipeline described in Section~\ref{sec:software} and the priors listed in Table~\ref{tab:priors} (we use the uniform prior on $\log_{10}S_0$ here).
We fix the WN parameters to their injected values and search for a CURN and an eccentric PTA signal while analytically marginalizing the {linearized} timing model.

The posterior samples obtained from these searches, {for the simulated dataset with injected luminosity distance = $D_\text{L}^\text{true}/5$}, are shown as corner plots in Figures~\ref{fig:sim_eonly} (E-only) and \ref{fig:sim_ep} (E+P).
In both E-only and E+P cases, we find that the recovered CURN parameter values are consistent with the injected values within the $2\sigma$ level.
We find that the estimated $A_\CURN$ and $\gamma_\CURN$ values in the E-only case show an offset as compared to the injected value whereas they are in closer agreement in the case of the E+P case.
Whether this is a general feature of E-only searches caused by the unmodeled pulsar terms will be investigated in a future work. 

The injected signal is detected in both E-only and E+P searches; the $\log_{10} S_0$ posterior distribution is consistent with the injected value although it has a long tail to the left.
The $l_0$, $\cos\iota$, and $\eta$ posterior distributions are broad but informative and consistent with the injected values.
The $e_0$ posterior distributions for both E-only and E+P searches are also informative and consistent with the injected values but are also consistent with $e=0$.
This is especially true in the case of the E+P search, where the posterior becomes flat as $e\rightarrow 0$.
Further, in both cases, $\varpi_0$ has a nearly flat posterior and $\psi$ has a bimodal posterior where one of the peaks is consistent with the injected value.
The bimodal structure in the posterior distribution of $\psi$ is a consequence of the $\pi$-periodicity of the PTA signal expression in this parameter (see equation~\eqref{eq:R(t)_px}).
Overall, we see that both the E-only and E+P searches are able to detect the injected signal but are not able to estimate many of the model parameters precisely.

{In the case of E+P search, we found that the posterior distributions for $D_p$, $\varpi_p$, and $l_p$ for all the pulsars are very similar to the prior distributions of the respective parameters, suggesting that the data does not provide any new insights or information about these parameters.}
Further, we find that the marginalized joint posterior distribution of $e_0$ and $\eta$ has a geometry reminiscent of the prior distribution plotted in Figure \ref{fig:validated-prior}.
To see if the $(e_0,\eta)$ joint posterior provides more information than the corresponding prior distribution, we overplot them in Figure \ref{fig:sim_e+p-prior-comp}. 
To provide a fair comparison, we restrict the prior samples in Figure \ref{fig:sim_e+p-prior-comp} to those where the $\log_{10} S_0$ values fall within the 95\% credible interval of $\log_{10} S_0$ derived from the posterior distribution.
It is clear from Figure \ref{fig:sim_e+p-prior-comp} that the upper limit on $e_0$ seen in Figure \ref{fig:sim_ep} is informed by the data, whereas the lower limit on $\eta$ is an artifact of the prior.
We have found that similar conclusions can be drawn for the posterior distributions for $e_0$ and $\eta$ for the E-only search.

We repeated the above-described simulation studies with different injected values of $D_{\rm L}$, {namely $D_\text{L}^\text{true}$ and $D_\text{L}^\text{true}/10$}.
Injecting the original value of the $D_{\rm L}$ ($D_\text{L}^\text{true}$) results in a non-detection where we are able to put an upper limit on $\log_{10} S_0$ in both E-only and E+P cases.
It turned out that the posterior distribution {in this case} does not gain any significant information from the data for certain combinations of $e_0$ and $\eta$, where the $\log_{10} S_0$ upper limit is not physically meaningful {(this is also seen in the $D_{\rm L}^{\rm true}/5$ case, as seen in Figure \ref{fig:sim_e+p-prior-comp})}.
We defer a detailed discussion on this caveat to Section \ref{sec:results}.
Injecting a lower value of the $D_{\rm L}$ (namely $D_\text{L}^\text{true}/10$) resulted in a detection in the E-only search, and the geometry of the posterior distribution turned out to be very similar to Figure \ref{fig:sim_eonly}.
Unfortunately, the corresponding E+P search failed to adequately explore the parameter space.
We suspect that this is due to a combination of its high dimensionality arising from the pulsar term parameters and complex geometry arising from the strong signal present in the data,
i.e., our search method is inadequate in the strong-signal regime in the E+P case.

The limited number of simulations we performed indicate that an E-only search can detect a PTA signal from an eccentric SMBHB, which is consistent with the results found in \citet{CharisiTaylor+2023} for circular binaries.
However, a more rigorous simulation study is required to {determine} whether an E-only search will introduce a bias in the parameter estimations (or the upper limits in case of a non-detection) compared to an E+P search for eccentric binaries in PTA data sets.

\section{Results}
\label{sec:results}

We now present the results of our search for continuous GWs from an eccentric SMBHB in 3C~66B in the NG12.5 data set.
We run both an E-only search and an E+P search using the uniform priors for $\log_{10}S_0$ for the detection analysis.
The full signal model including a linearized timing model, WN, IRN, CURN, and eccentric PTA signal, as described in equation~\eqref{eq:full_signal_model}, is used.
The Bayes factors estimated using the Savage-Dickey formula (see Subsection \ref{sec:model-comp}) turned out to be $0.99\pm 0.01$ for the E-only model and $1.01\pm 0.01$ for the E+P model.
It is evident that we did not detect any GWs from an eccentric SMBHB in 3C~66B as these numbers do not favor the presence of an eccentric PTA signal in our data.

In the absence of a detection, we now turn our attention to constraining the PTA signal amplitude $S_0$ and the chirp mass $\Mch$ for a possible eccentric SMBHB present in 3C~66B.
The posterior samples for the upper limit analysis are obtained by reweighting the posterior samples from the detection analysis as described in \citet{HourihaneMeyers+2022}.
Before calculating the upper limits, let us recall that, for certain parts of the parameter space, the posterior distribution of $S_0$ may only be informed by the prior distribution, determined in part by the validation criteria described in Subsection \ref{sec:priors} and visualized in Figure \ref{fig:validated-prior} 
(this is seen to some extent in the injection study results shown in Figures \ref{fig:sim_ep} and \ref{fig:sim_e+p-prior-comp}).
Therefore, the upper limit on $S_0$ (and $\Mch$) calculated for those parts of the parameter space can be misleading.
Since $e_0$, $\eta$, and $\log_{10}S_0$ are not independent parameters in the prior distribution, it is more meaningful to explore the posterior distributions of $\log_{10}S_0$ as functions of $(e_0,\eta)$ rather than marginalize over these parameters.

\begin{figure*}
    \centering
    \begin{tabular}{c}
    (a)\cr \includegraphics[width=0.65\textwidth]{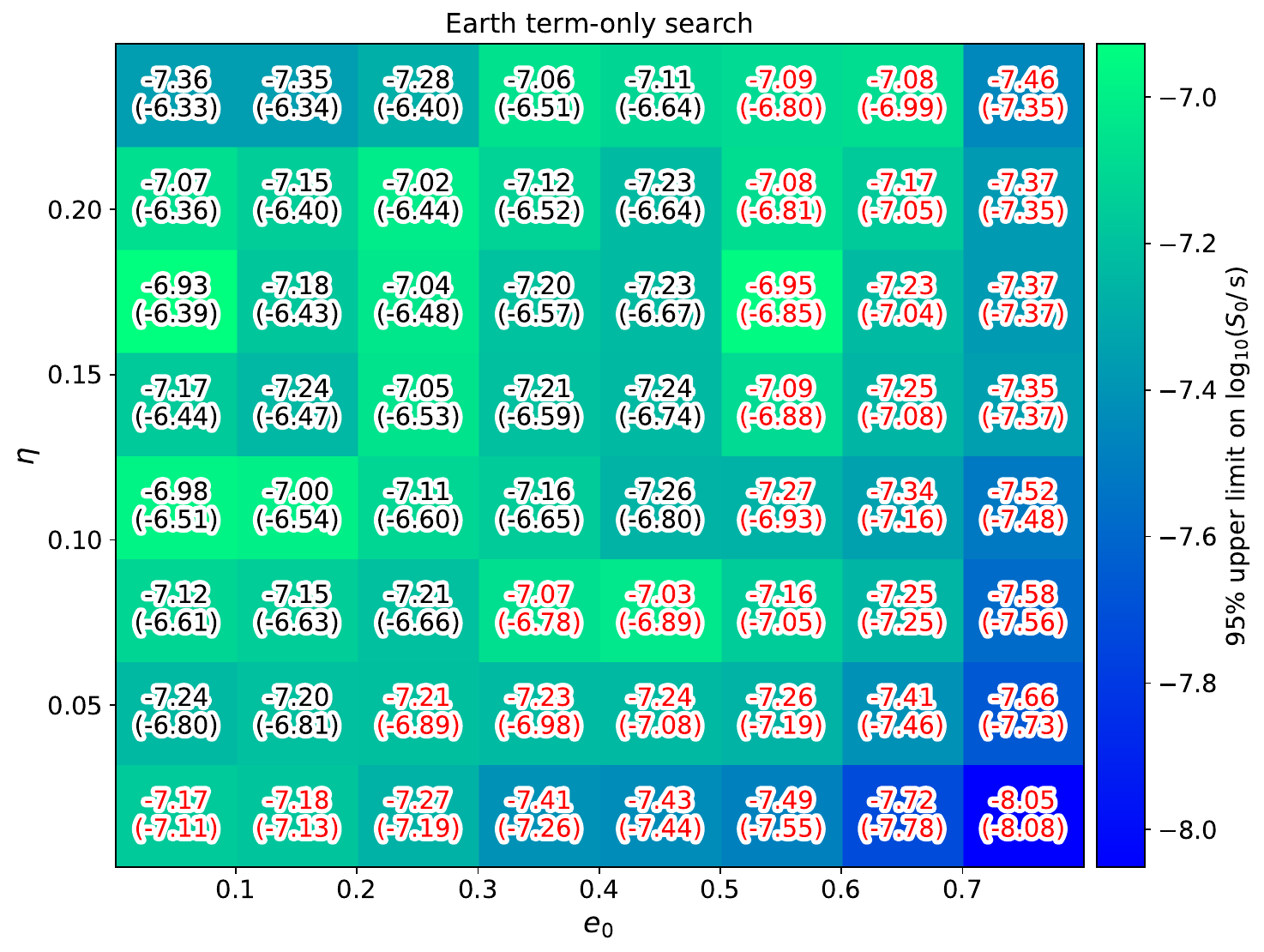}  \\
    (b)\cr \includegraphics[width=0.65\textwidth]{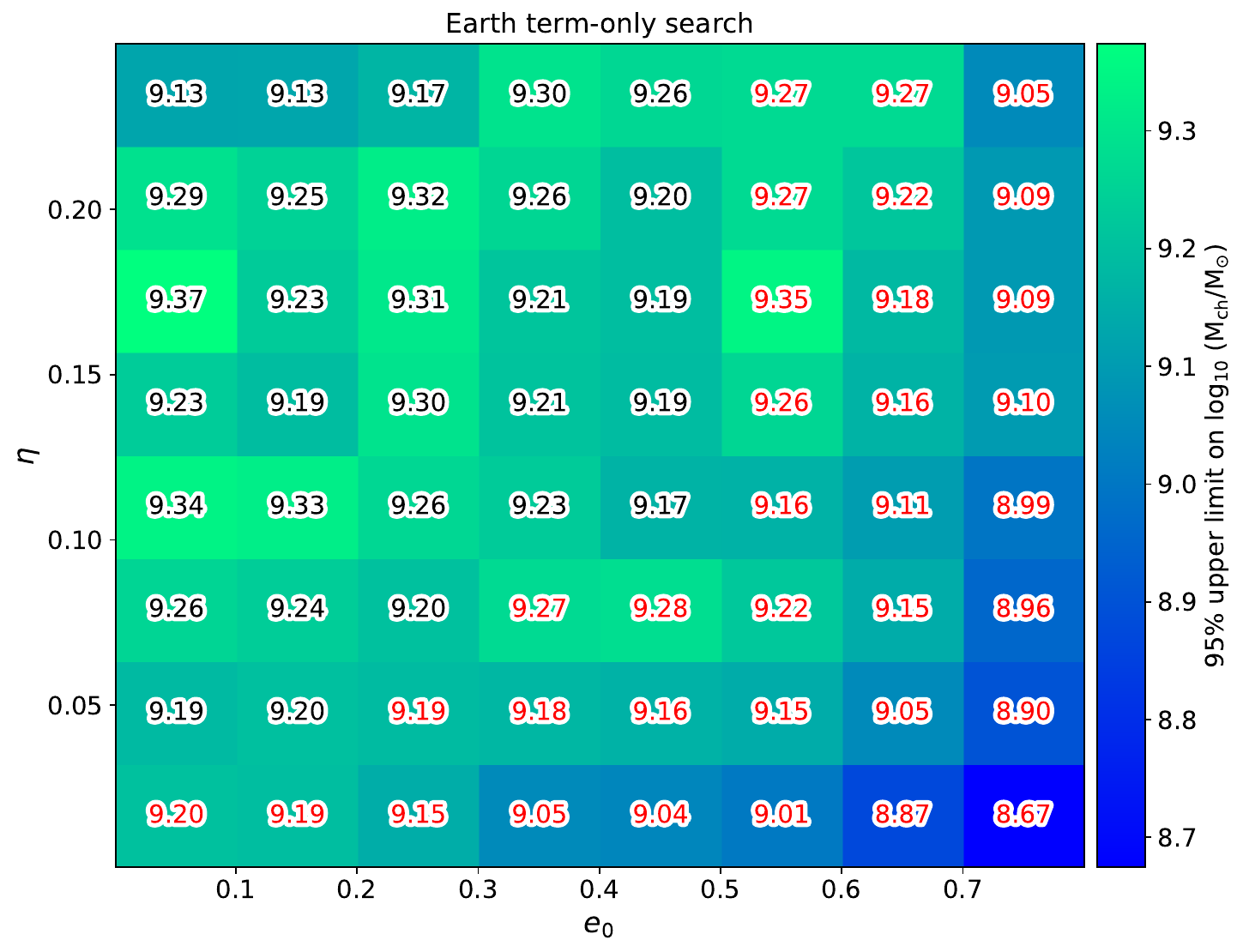}  
    \end{tabular}
    \caption{95\%  upper limits on $\log_{10}S_0$ (panel a, top) and $\log_{10}\Mch$ (panel b, bottom) obtained from the Earth term-only search, binned in $e_0$ and $\eta$, plotted using color maps.
    $S_0$ and $\Mch$ are expressed in units of seconds and $\Msun$, respectively.
    In each $(e_0,\eta)$ pixel of each panel, the 95\%  upper limit value is quoted.
    In the top panel, the 95\%  upper limit obtained from the prior distribution is quoted in parentheses.
    The pixels where the $\log_{10}S_0$ upper limit may be restricted by the prior distribution and not determined by the data are highlighted with red font color and do not represent physically relevant results (they arise due to the limitations of our PTA signal model, see the text for discussion).
    We are able to obtain astrophysically meaningful constraints for $e_0<0.5$ and $\eta>0.1$.
    }
    \label{fig:earth_S0_95}
\end{figure*}

\begin{figure*}
    \centering
    \begin{tabular}{c}
    (a)\cr \includegraphics[width=0.65\textwidth]{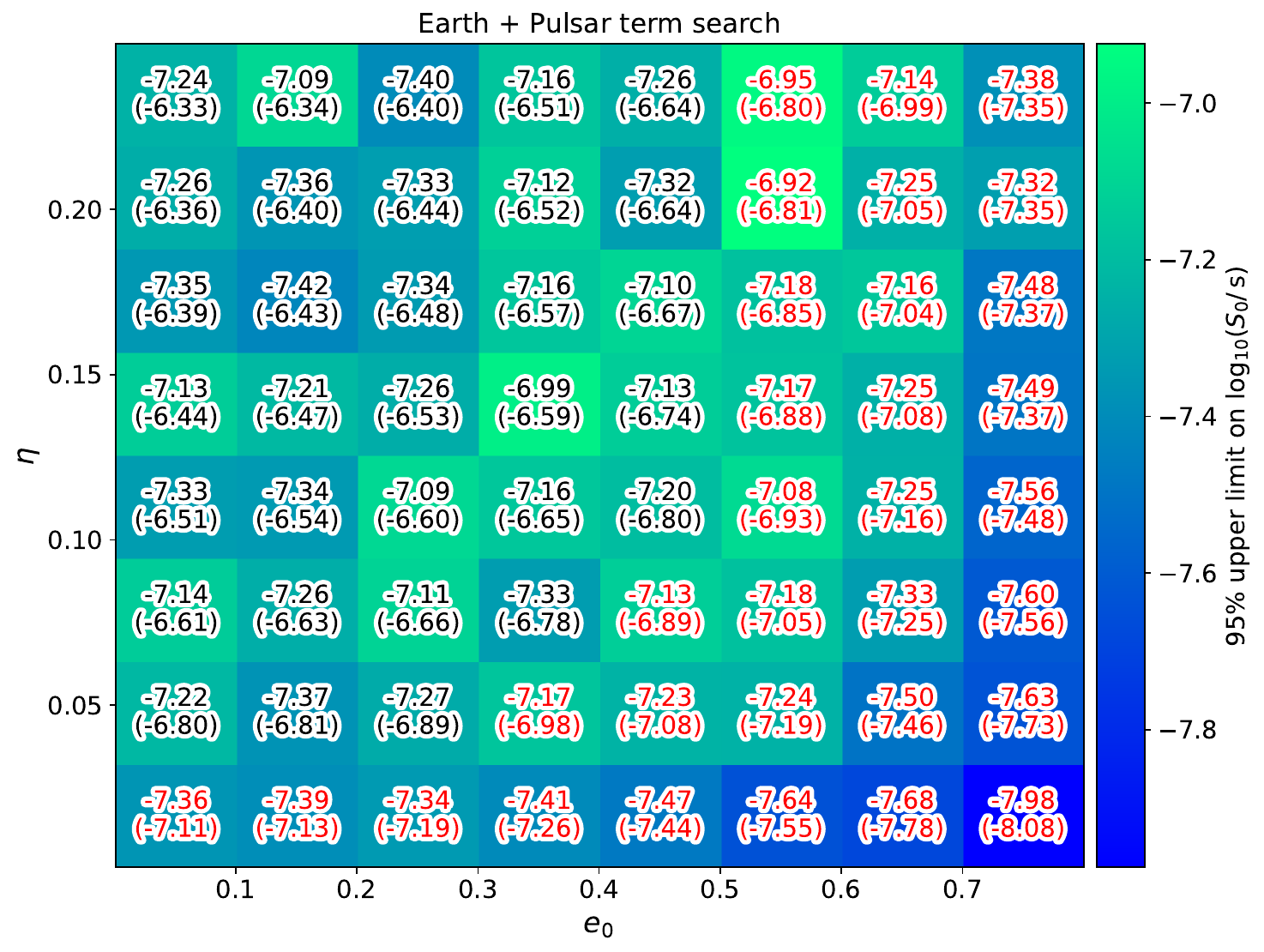}  \\
    (b)\cr \includegraphics[width=0.65\textwidth]{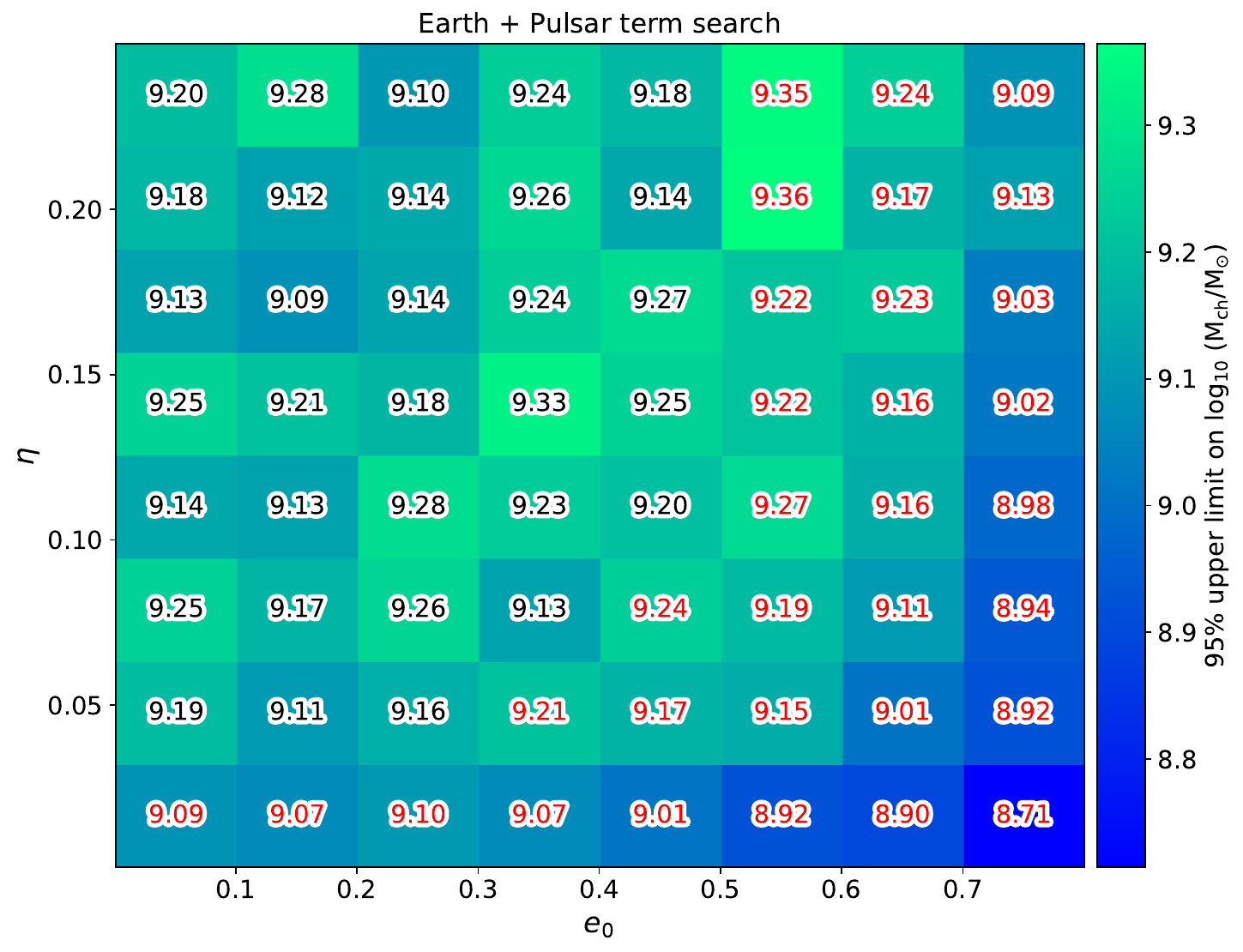}  
    \end{tabular}
    \caption{95\%  upper limits on $\log_{10}S_0$ (panel a, top) and $\log_{10}\Mch$ (panel b, bottom) obtained from the Earth+Pulsar term search, binned in $e_0$ and $\eta$, plotted using color maps.
    The plotting conventions are identical to those in Figure \ref{fig:earth_S0_95}.
    We are able to obtain physically relevant constraints for $e_0<0.5$ and $\eta>0.1$.
    For comparison, the 95\% upper limit from the Earth+Pulsar term targeted search of a circular binary in 3C~66B using NG12.5 data set is $\log_{10}\Mch < 9.15$ \citep{ArzoumanianBaker+2023}.}
    \label{fig:earth_pulsar_S0_95}
\end{figure*}

We divide the prior ranges for both $e_0$ and $\eta$ into 8 equally spaced bins and calculate the upper limit on $S_0$ for each $e_0$ and $\eta$ bin from the posterior distribution while marginalized over all other parameters.
The 95\%  upper limits on $\log_{10}S_0$ for all pairs of bins from the E-only search are plotted using color maps in the upper panel of Figure~\ref{fig:earth_S0_95}.
Recall that the upper limits are obtained after reweighting the samples generated by the detection analysis run.
Further, we calculate the 95\% upper limit on $\log_{10}S_0$ from the modified joint prior distributions (Figure~\ref{fig:validated-prior}) for each $(e_0,\eta)$ pixel following the same way.
The values of the 95\%  upper limits on $\log_{10}S_0$ derived from the posterior distribution are quoted in each $(e_0,\eta)$ pixel along with the corresponding 95\% upper limits obtained from the prior distribution within parentheses. 
We also plot the 95\%  upper limit on $\log_{10}\Mch$ for the E-only search in the lower panel of Figure~\ref{fig:earth_S0_95} and the corresponding values for each pixel are also quoted.
We see from panel (a) of Figure~\ref{fig:earth_S0_95} that for pixels with high $e_0$ and/or low $\eta$ values, the 95\% upper limits obtained from the posterior distribution of $\log_{10}S_0$ are very close to those obtained from the prior.
This is an indication that the upper limits of $\log_{10} S_0$ for those pixels is determined by the prior, which we adopted to avoid regions of parameter space where the signal model is not valid, rather than the data.
Hence, our upper limits for such pixels will not be astrophysically meaningful due to being affected by the limitations of our PTA signal model.
We discern the pixels where the upper limit may be arising due to the restrictions imposed by our joint prior distribution and not by the data, using the following criterion
\begin{equation}
\frac{\log_{10}S_0^{\text{post}}[95\%] - \log_{10}S_0^{\text{prior}}[95\%]}{\log_{10}S_0^{\text{prior}}[95\%]} < 0.05\,,
\end{equation}
where $\log_{10}S_0^{\text{prior}}[95\%]$ and $\log_{10}S_0^{\text{post}}[95\%]$ represent the 95\%  upper limits obtained from the prior and posterior distributions, respectively.
This implies that we treat the upper limit for a pixel as \textit{invalid} if the fractional difference between $\log_{10}S_0^{\text{prior}}[95\%]$ and $\log_{10}S_0^{\text{post}}[95\%]$ is less than a conservative tolerance of 5\%.
These upper limits do not represent physically meaningful results, arising due to the limitations of our PTA signal model, and we highlight such pixels using red font color in Figure~\ref{fig:earth_S0_95}.
Note that, in principle, the upper limits can be limited by the prior distribution even in cases where the posterior distribution gains significant information over the prior from the data.
This is why we base the above criterion on the upper limits rather than on a statistic such as Kullback–Leibler divergence \citep{KullbackLeibler1951} that treats the prior and posterior distributions more holistically.

\begin{figure*}
    \centering
    \includegraphics[width=1\textwidth]{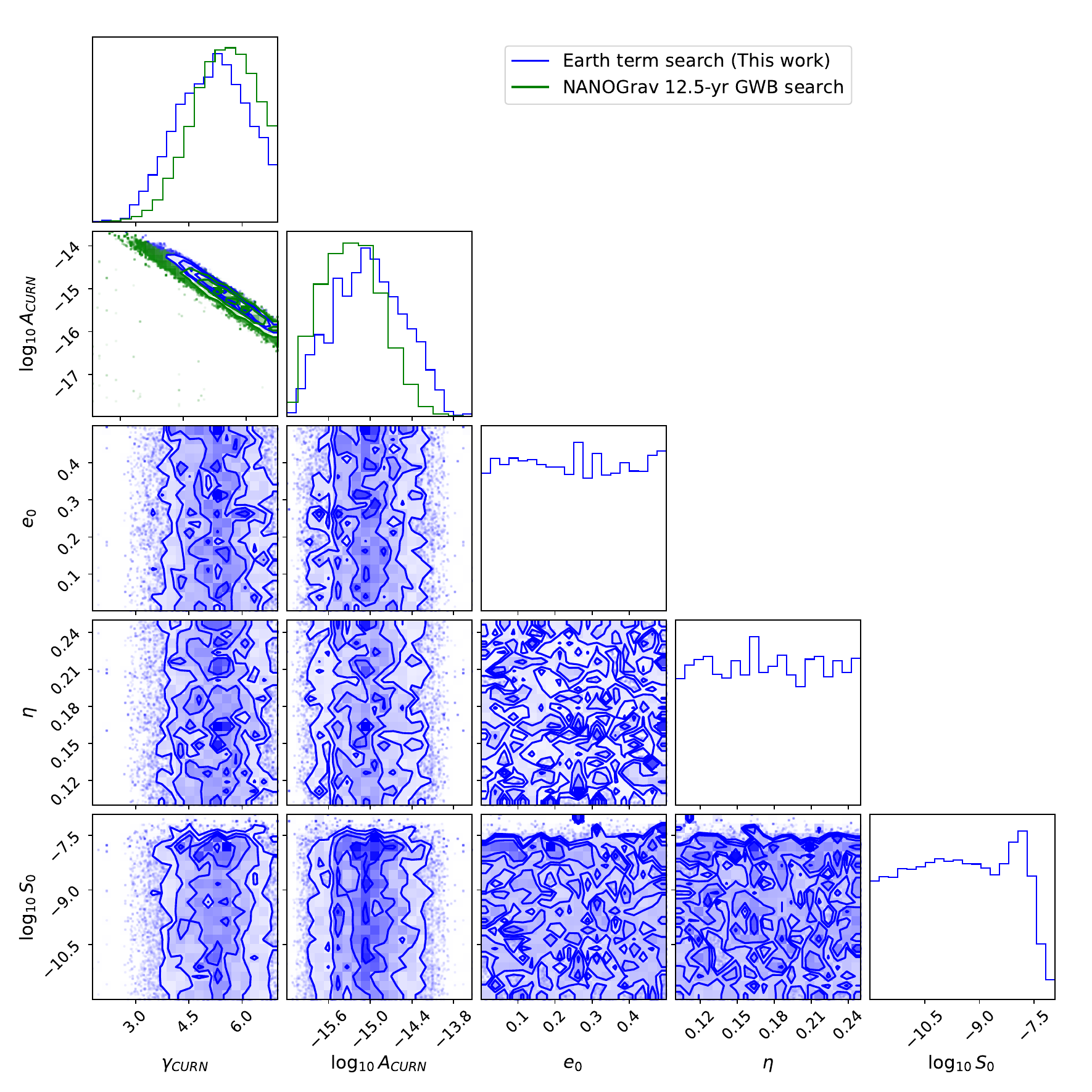}
    \caption{Posterior distribution for the CURN parameters ($\gamma_\text{CURN}$ and $\log_{10} A_\text{CURN}$) and the eccentric SMBHB signal parameters ($e_0$, $\eta$, and $\log_{10} S_0$) for the Earth term-only search marginalized over all other parameters (plotted in blue).
    We only include samples with $e<0.5$ and $\eta>0.1$ in this plot as the posterior distribution is significantly affected by the prior outside this region of the parameter space.
    The CURN parameters obtained from the NANOGrav 12.5-year GWB search are overplotted in green for comparison.
   }
    \label{fig:gwecc_posterior_E}
\end{figure*}

\begin{figure*}
    \centering
    \includegraphics[width=1\textwidth]{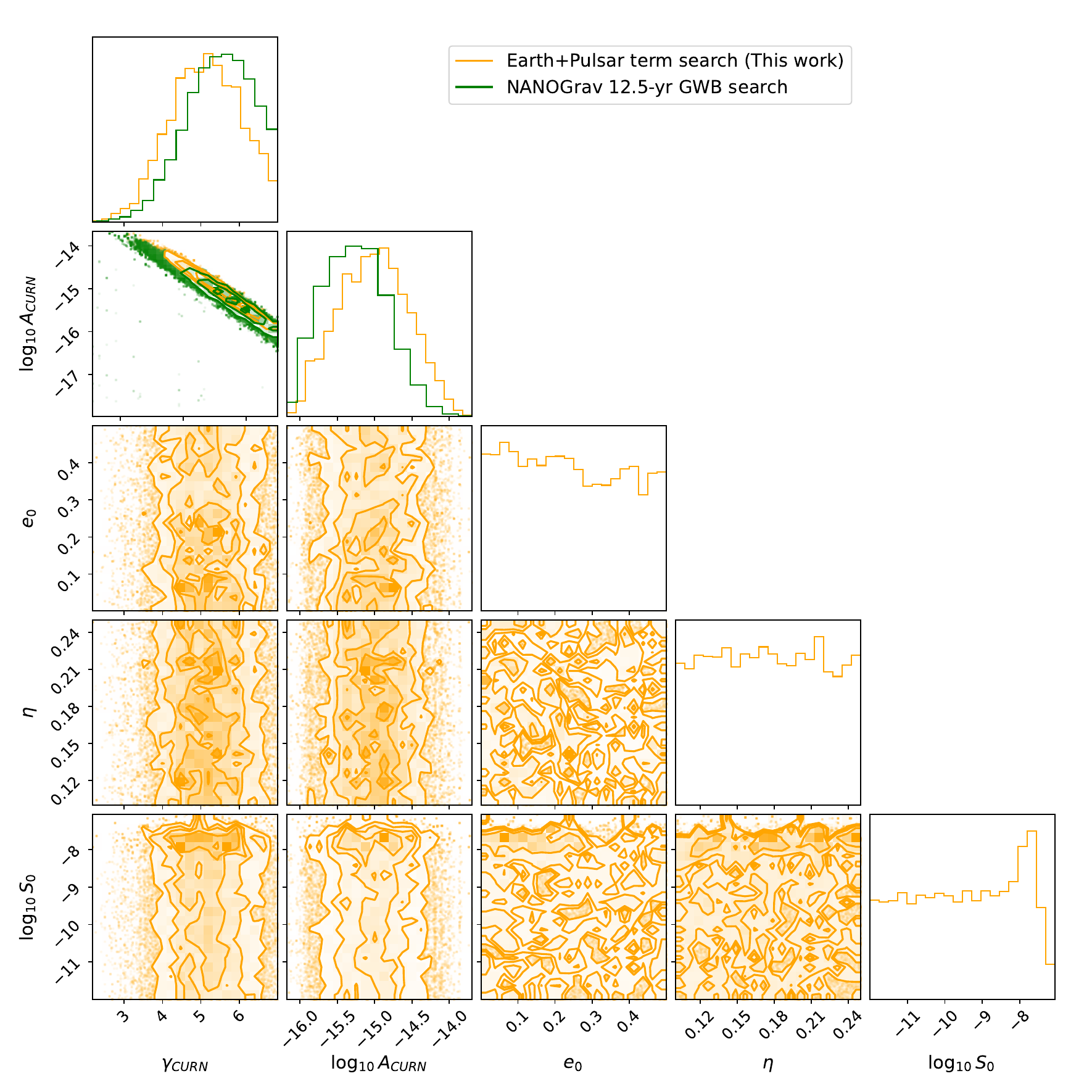}
    \caption{Posterior distribution for the CURN parameters ($\gamma_\text{CURN}$ and $\log_{10} A_\text{CURN}$) and the eccentric SMBHB signal parameters ($e_0$, $\eta$, and $\log_{10} S_0$) for the Earth+Pulsar term search marginalized over all other parameters (plotted in orange).
    We only include samples with $e<0.5$ and $\eta>0.1$ in this plot as the posterior distribution is significantly affected by the prior outside this region of the parameter space.
    The CURN parameters obtained from the NANOGrav 12.5-year GWB search are overplotted in green for comparison.
    }
    \label{fig:gwecc_posterior_EP}
\end{figure*}

\begin{figure*}
    \centering
    \includegraphics[width=0.7\textwidth]{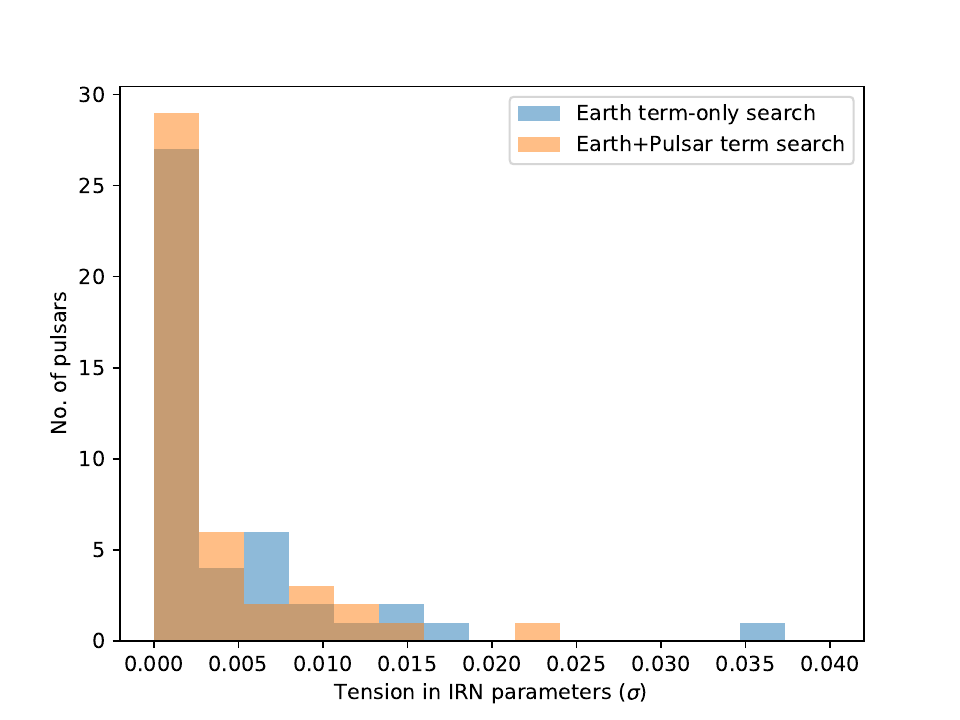}
    \caption{The Raveri-Doux tension between marginalized IRN posterior distributions obtained from the NANOGrav 12.5-year GWB search and this work for different pulsars, plotted as a histogram.
    The blue histogram corresponds to the Earth term-only search and the orange histogram corresponds to the Earth+Pulsar term search.
    The tension remains less than $0.04\sigma$ for all pulsars in both searches.
    This implies that our searches are not significantly affected by the IRN.
    }
    \label{fig:psr-noise-tension}
\end{figure*}

Upper limits on $\log_{10}S_0$ and $\log_{10}\Mch$ for the Earth+Pulsar term search are calculated in same way and shown in Figure~\ref{fig:earth_pulsar_S0_95}.
It can be seen from Figures \ref{fig:earth_S0_95} and \ref{fig:earth_pulsar_S0_95} that, for the subset of the parameter space where $e<0.5$ and $\eta>0.1$, the upper limits are informed by the data and not just the restrictions of the prior.
This is consistent with the shape of the joint prior distribution seen in Figure \ref{fig:validated-prior}.
We have also computed the Savage-Dickey Bayes factor in each pixel, and they remain close to 1 indicating non-detections in both E-only and E+P searches.

We plot the marginalized posterior distributions for the CURN parameters and $\log_{10} S_0$, $e_0$, and $\eta$ obtained from the E-only and E+P detection analyses in Figures \ref{fig:gwecc_posterior_E} and \ref{fig:gwecc_posterior_EP} respectively.
We omit the samples with $e_0>0.5$ or $\eta<0.1$ in these plots since they are significantly affected by the prior distribution as seen above.
The 95\%  upper limits (calculated after reweighting) on $S_0$ are $88.1 \pm 3.7$ ns and $81.74 \pm 0.86$ ns respectively for E-only and E+P analyses while restricting  $e_0<0.5$ and $\eta>0.1$.
Similar 95\%  upper limits on $\Mch$ are $(1.98 \pm 0.05) \times 10^9 \,\Msun$ and $(1.89 \pm 0.01) \times 10^9 \, \Msun$.
The uncertainties in the upper limits are calculated using the bootstrap method.
We also overplot the posterior distributions for the CURN parameters obtained from the NG12.5 GWB search \citep{NG2020_12yr_GWB} in Figures \ref{fig:gwecc_posterior_E} and \ref{fig:gwecc_posterior_EP}.
We observe that the addition of the eccentric SMBHB signal does not alter the posterior distribution of the CURN parameters appreciably in both E-only and E+P searches.
This indicates the robustness of our search against the leakage of power between the CURN and the eccentric SMBHB signal. 
{Moreover, for the E+P search, we found that the posterior distributions for $D_p$, $\varpi_p$, and $l_p$ for all the pulsars closely align with their respective prior distributions.}

We now turn our attention to the robustness of our search against the leakage of power between IRN and the eccentric SMBHB signal.
We do this by comparing the posterior distributions of the IRN parameters obtained from our E-only and E+P detection analyses against those obtained from the NG12.5 GWB search using the Raveri-Doux tension statistic, available in the \texttt{tensiometer} package \citep{RaveriDoux2021}.
Figure \ref{fig:psr-noise-tension} shows a histogram of the tensiometer statistics computed for different pulsars, and we see that the IRN parameter posteriors from our analyses agree with those from the NG12.5 GWB search \citep{NG2020_12yr_GWB} within the $ 0.04\sigma$ level for all pulsars.
This gives us confidence that the search for the eccentric SMBHB signal is not affected significantly by the IRN.

\section{Summary and Discussion}
\label{sec:discussion}

{We have developed a pipeline for performing a targeted search for continuous GWs from individual eccentric SMBHB in a PTA data set.
We calculate the pulsar timing residuals induced by GWs from an eccentric SMBHB using the semi-analytic approach presented in \citet{SusobhananGopakumar+2020} and \citet{Susobhanan2023}.
We tested our pipeline by applying it to simulated data sets, and we have performed an Earth term-only (E-only) search where contributions from the pulsar term are neglected and an Earth+Pulsar term (E+P) search where contributions from both terms are included.
In both cases, we have found that the marginalized posterior distributions for our free model parameters are consistent with the injected values for the simulated data.}
Thereafter, we performed both E-only and E+P targeted searches for GWs from an eccentric SMBHB in the radio galaxy 3C~66B in the NANOGrav 12.5-year data set.
In addition to the eccentric SMBHB signal, our model incorporates a CURN process, detected as a precursor to the GWB in the NG12.5 data set \citep{NG2020_12yr_GWB}, IRN processes for each pulsar, and WN processes for each pulsar.
We fix the WN parameters for each pulsar to the values obtained from its single-pulsar noise analysis.
This is the first time a multi-messenger targeted search for GWs from an eccentric SMBHB is performed on a full-scale PTA data set.

We did not find any evidence for continuous GWs from an eccentric SMBHB in 3C~66B in our search.
Hence, we calculated the upper limits on the GW signal amplitude ($S_0$) and the chirp mass ($\Mch$) of the SMBHB candidate in 3C~66B as a function of the initial eccentricity ($e_0$) and symmetric mass ratio ($\eta$) of the binary.
We found that for certain combinations for $e_0$ and $\eta$ of the binary, the $\log_{10}S_0$ upper limit is limited by the prior and is not informed by the data.
This arises due to the limitations of our PTA signal model and such upper limits should not be considered astrophysically meaningful.
For the parameter range of $e_0 < 0.5$ and $\eta > 0.1$, the upper limits are not influenced by the limitations of our PTA signal model and are informed by the data.
In that regime, the 95\%  upper limits obtained from the E-only search are $88.1 \pm 3.7$ ns for $S_0$ and $(1.98 \pm 0.05) \times 10^9 \Msun$ for $\Mch$.
Similar 95\%  upper limits obtained from the E+P search are $81.74 \pm 0.86$ ns for $S_0$ and $(1.89 \pm 0.01) \times 10^9 \Msun$ for $\Mch$.
We note that the peculiar motion of 3C~66B can give rise to an uncertainty in the luminosity distance estimate from the redshift of that galaxy.
This may lead to a systematic bias in the estimated upper limits on $\Mch$ whereas the upper limits on $S_0$ should be unaffected.
We also found that the posterior distributions of the CURN and IRN parameters in both our searches are consistent with those obtained from the NG12.5 GWB search  \citep{NG2020_12yr_GWB}.

The SMBHB model of 3C~66B proposed by \citet{IguchiOkudaSudou2010} with a chirp mass of $7.9^{+3.8}_{-4.5} \times 10^8 \Msun$ is consistent with the upper limits we obtain from our analysis.
We also see that the upper limits on the chirp mass from our targeted search for GWs from eccentric binary in 3C~66B are higher than that obtained from the targeted search from circular binaries in NG12.5 data set \citep{ArzoumanianBaker+2023}.
This may be because of the higher number of free parameters present in our eccentric search as compared to the circular search. 
Another reason for this may be that, as the eccentricity pushes the signal power to higher frequencies, the dominant frequency might lie at high frequencies where the data is less sensitive due to white noise.
Further, the power of an eccentric signal is spread across multiple frequencies and the signal at some of these frequencies may lie below the noise threshold.

\begin{figure*}
    \centering    \includegraphics[width=0.6\textwidth]{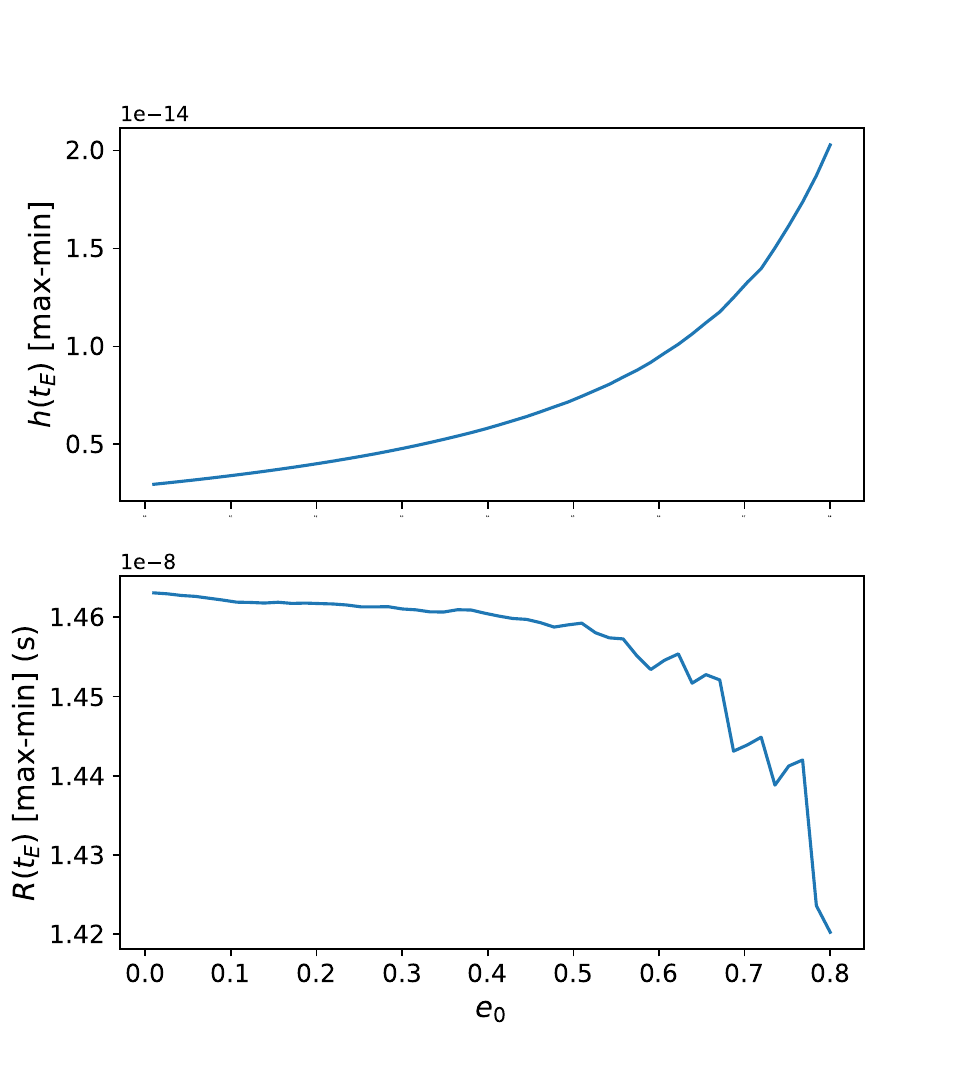}
    \caption{Maximum$-$minimum (crest-to-trough) amplitude of the gravitational waveform (top panel) and PTA signal (bottom panel) for PSR J1909$-$3744 due to an SMBHB as a function of orbital eccentricity.
    The crest-to-trough amplitude is the difference between the maximum and minimum values of the waveform for a given time span and is not to be confused with $S_0$.
    We have only considered the Earth term.
    The masses of the binary are fixed to the 3C~66B binary model values given in \citet{IguchiOkudaSudou2010}: $m_1 = 1.2\times10^9 \, \Msun$, $m_2 = 7.0\times10^8 \, \Msun$.
    We have used $\cos\iota = 1$, $\psi = 0$, $l_0=0$, and $\gamma_0=0$, and a time span of 30 years.
    The sky position and luminosity distance of the binary are taken from Table~\ref{tab:priors}.
    We do not find any strong variation in the PTA signal amplitude with eccentricity, whereas the waveform amplitude is an increasing function of eccentricity.
    Note that the overall weak decreasing trend seen in the bottom plot is not universal and changes depending on binary parameter values.}
    \label{fig:S0minmax_e0}
\end{figure*}

We do not find any clear trend in the upper limits on $\log_{10}S_0$ as a function of $e_0$ or $\eta$ for both E-only and E+P searches, as evident from the 2-D posterior distributions shown in Figures~\ref{fig:gwecc_posterior_E} and \ref{fig:gwecc_posterior_EP}.
This is not surprising because of the parameterization used for our targeted search (discussed in Section~\ref{sec:targeted-search-params}), where the overall PTA signal amplitude at the Newtonian order is just given by $S_0$.
Although different values of $e_0$ and $\eta$ can lead to different temporal evolution of the PTA signal amplitude $S(t)$ through $\varsigma(t)$ (due to different evolution of $n(t)$ and $k(t)$ appearing at higher post-Newtonian order), we suspect that for a non-detection these do not affect the upper limits significantly.
This is also evident from Figure~\ref{fig:S0minmax_e0}, where we show the crest-to-trough amplitudes of the gravitational waveform (top panel) and the PTA signal (not to be confused with $S_0$) induced by GWs from an SMBHB (bottom panel) as a function of the eccentricity when we keep all the other binary parameters fixed.
We see that although the gravitational waveform amplitude increases with the eccentricity, the amplitude of the PTA signal does not vary significantly with the eccentricity.
Therefore, we expect that the upper limit on $S_0$ should also not depend significantly on the eccentricity of the binary.
Similarly, upper limits on $\Mch$ also should not depend significantly on $e_0$ and $\eta$ as we can see from equation~\eqref{eq:S0_Mch_relation} that at the Newtonian order, the PTA signal amplitude depends only on $\Mch$ and not on $e_0$ or $\eta$.
Please note that Figure 10 does not imply anything about the measurability of eccentricity in the case of a detection, which would only depend on the quality and sensitivity of the dataset as well as the strength of the signal.

In this work, we have used a Gaussian red noise process with a 30-component power-law spectrum for modeling individual pulsar red noise (IRN) and a DMX model to account for dispersion measure (DM) variations, following standard NANOGrav practices \citep{NG2020_12yr_GWB, NG2023_15yr_GWB, ArzoumanianBaker+2023}.
However, different PTAs make different choices for modeling IRN and DM variations in their data.
Recently, \citet{IPTA_3P+_2023} compared the GWB search results from different PTAs and showed that, despite making different modeling choices, there is no significant difference in the GWB parameters measured by different PTAs.
Further, for the majority of the pulsars, the IRN parameters are consistent among different PTAs.
This suggests that the choice of individual pulsar noise models should not significantly affect the upper limits obtained in our analysis.

We find that certain aspects of our methods and the search pipeline can be improved further, especially in order to perform targeted PTA searches for GWs from eccentric binaries based on a catalog of SMBHB candidates.
For many such candidates, the orbital period of the binary is not very accurately known, and therefore using a Gaussian prior for the initial binary orbital period ($P_0$) around the proposed values instead of fixing it should lead to more realistic results.
Further, due to the limitation of our PTA signal model, we were able to calculate meaningful upper limits only for a certain regime of the parameter space (of $e_0$ and $\eta$) and that regime depends on the reference time $t_0$ and the total span of our data.
A later reference time $t_0$ for the binary parameters will extend that regime whereas a longer data span will shrink it.
Therefore, mitigating the limitations of our PTA signal model will be crucial for exploring the whole $(e_0,\eta)$ parameter space and obtaining meaningful upper limits on the PTA signal amplitude.
This will be especially relevant for CHIME, FAST, and SKA-era PTA data sets, where a high observation cadence will make the data sensitive to higher GW frequencies, which correspond to highly relativistic SMBHBs.
Ways to extend the validity of our PTA signal model closer to the merger event include the effective one-body formalism \citep{HindererBabak2017} and replacing the currently unmodeled merger and ringdown parts of the waveform with a generic burst signal model.

A new method to search for individual sources in PTA data sets, named \texttt{QuickCW} \citep{BecsyCornishDigman2022}, was introduced recently and was applied to great effect to search for circular SMBHBs in the NANOGrav 15-year data set \citep{NG2023_15yr_individual}.
\texttt{QuickCW} exploits the mathematical structure of the PTA signal expression to accelerate the likelihood computation in the case of projection parameter updates (e.g., in our model, the projection parameters are $S_0$, $\psi$, $\cos\iota$, $\varpi_0$, and $\varpi_p$ for each pulsar).
Extending this method for eccentric SMBHB searches will be a promising avenue to explore in order to keep future searches computationally feasible in the face of growing data volumes.
Further, \citet{CharisiTaylor+2023} showed that ignoring the pulsar term contributions does not significantly affect the parameter estimation in the case of targeted PTA searches for circular SMBHBs.
Performing detailed simulation studies to confirm that the same is true for targeted searches of eccentric SMBHBs for both detection and upper limit analyses will help us efficiently perform such searches in future PTA data sets by ignoring the pulsar term.
It will also be interesting to explore the possibility of using Hamiltonian Monte Carlo (HMC) to search for SMBHB signals in PTA data sets, given its performance advantages over other types of MCMC methods in high-dimensional parameter spaces (see \citet{FreedmanJohnson+2023} for a PTA application of HMC).
These considerations will be especially relevant for conducting an eccentric SMBHB search in the upcoming IPTA Data Release 3, which is expected to be much larger and more sensitive than the NG12.5 data set.

\section*{Data availability}
The NANOGrav 12.5-year data set is available at \url{https://nanograv.org/science/data/125-year-pulsar-timing-array-data-release}, and the MCMC chains from the NANOGrav 12.5-year GWB search are available at \url{https://nanograv.org/science/data/125-year-stochastic-gravitational-wave-background-search}.

\section*{Software}
\texttt{ENTERPRISE} \citep{EllisVallisneri+2017,JohnsonMeyers+2023},
\texttt{enterprise\_extensions} \citep{TaylorBaker+2021,JohnsonMeyers+2023},
\texttt{PTMCMCSampler} \citep{EllisvanHaasteren2017,JohnsonMeyers+2023},
\texttt{GWecc.jl} \citep{Susobhanan2023},
\texttt{libstempo} \citep{Vallisneri2020},
\texttt{numpy} \citep{HarrisMillman+2020}, 
\texttt{pandas} \citep{McKinney2010}, 
\texttt{wquantiles} \citep{Sabater2015},  
\texttt{tensiometer} \citep{RaveriDoux2021},
\texttt{matplotlib} \citep{Hunter2007}, 
\texttt{corner} \citep{Foreman-Mackey2016}.

\section*{Author contributions}
This paper uses an alphabetical-order author list to recognize that a large, decade-timescale project such as NANOGrav is necessarily the result of the work of many people. 
All authors contributed to the activities of the NANOGrav collaboration leading to the work presented here and reviewed the manuscript, text, and figures before the paper’s submission. 
Additional specific contributions to this paper are as follows.
The NANOGrav 12.5-year targeted search for an eccentric SMBHB source in 3C 66B was led by LD with crucial inputs from AS and SBS. 
BDC led this project during its initial exploratory stages with inputs from CAW, SBS, LD, AS, and AG. 
The theoretical aspects of this project were investigated by LD and AS. 
The data analysis and visualization scripts used in this project were developed by LD, AS, and BDC with inputs from CAW. 
The \gwecc{} package is developed and maintained by AS. 
This manuscript was prepared by AS and LD with inputs from SBS, AG, DLK, MAM, SJV, CAW, BB, MC, and TD.
ZA, HB, PRB, HTC, MED, PBD, TD, JAE, RDF, ECF, EF, NG-D, PAG, DCG, MLJ, MTL, DRL, RSL, JL, MAM, CN, DJN, TTP, NSP, SMR, KS, IHS, RS, JKS, RS, and SJV developed the 12.5-year data set through a combination of observations, arrival time calculations, data checks and refinements, and timing model development and analysis; additional specific contributions to the data set are summarized in \citet{AlamArzoumanian+2020}. 

~

\section*{Acknowledgements}

This work has been carried out by the NANOGrav collaboration, which receives support from the National Science Foundation (NSF) Physics Frontiers Center award numbers 1430284 and 2020265.
The Arecibo Observatory is a facility of the NSF operated under a cooperative agreement (No. AST-1744119) by the University of Central Florida (UCF) in alliance with Universidad Ana G. M{\'e}ndez (UAGM) and Yang Enterprises (YEI), Inc. 
The Green Bank Observatory is a facility of the NSF operated under a cooperative agreement by Associated Universities, Inc. 
The National Radio Astronomy Observatory is a facility of the NSF operated under a cooperative agreement by Associated Universities, Inc.
We thank the IPTA steering committee for their valuable inputs on the manuscript.
L.D. and A.S. thank Subhajit Dandapat for the useful discussions.
L.D. acknowledges the use of the Thorny Flat Supercomputing System at West Virginia University which is partly funded by the National Science Foundation (NSF) Major Research Instrumentation Program (MRI) Award \#1726534.
L.B. acknowledges support from the National Science Foundation under award AST-1909933 and from the Research Corporation for Science Advancement under Cottrell Scholar Award No. 27553.
P.R.B. is supported by the Science and Technology Facilities Council, grant number ST/W000946/1.
S.B. gratefully acknowledges the support of a Sloan Fellowship, and the support of NSF under award \#1815664.
M.C. and S.R.T. acknowledge support from NSF AST-2007993.
M.C. and N.S.P. were supported by the Vanderbilt Initiative in Data Intensive Astrophysics (VIDA) Fellowship.
Support for this work was provided by the NSF through the Grote Reber Fellowship Program administered by Associated Universities, Inc./National Radio Astronomy Observatory.
Support for H.T.C. is provided by NASA through the NASA Hubble Fellowship Program grant \#HST-HF2-51453.001 awarded by the Space Telescope Science Institute, which is operated by the Association of Universities for Research in Astronomy, Inc., for NASA, under contract NAS5-26555.
M.E.D. acknowledges support from the Naval Research Laboratory by NASA under contract S-15633Y.
T.D. and M.T.L. are supported by an NSF Astronomy and Astrophysics Grant (AAG) award number 2009468.
E.C.F. is supported by NASA under award number 80GSFC21M0002.
G.E.F., S.C.S., and S.J.V. are supported by NSF award PHY-2011772.
The Flatiron Institute is supported by the Simons Foundation.
A.D.J. and M.V. acknowledge support from the Caltech and Jet Propulsion Laboratory President's and Director's Research and Development Fund.
A.D.J. acknowledges support from the Sloan Foundation.
The work of N.La. and X.S. is partly supported by the George and Hannah Bolinger Memorial Fund in the College of Science at Oregon State University.
N.La. acknowledges the support from Larry W. Martin and Joyce B. O'Neill Endowed Fellowship in the College of Science at Oregon State University.
Part of this research was carried out at the Jet Propulsion Laboratory, California Institute of Technology, under a contract with the National Aeronautics and Space Administration (80NM0018D0004).
D.R.L. and M.A.M. are supported by NSF \#1458952.
M.A.M. is supported by NSF \#2009425.
C.M.F.M. was supported in part by the National Science Foundation under Grants No. NSF PHY-1748958 and AST-2106552.
A.Mi. is supported by the Deutsche Forschungsgemeinschaft under Germany's Excellence Strategy - EXC 2121 Quantum Universe - 390833306.
The Dunlap Institute is funded by an endowment established by the David Dunlap family and the University of Toronto.
K.D.O. was supported in part by NSF Grant No. 2207267.
T.T.P. acknowledges support from the Extragalactic Astrophysics Research Group at Eötvös Loránd University, funded by the Eötvös Loránd Research Network (ELKH), which was used during the development of this research.
S.M.R. and I.H.S. are CIFAR Fellows.
Portions of this work performed at NRL were supported by ONR 6.1 basic research funding.
J.D.R. also acknowledges support from start-up funds from Texas Tech University.
J.S. is supported by an NSF Astronomy and Astrophysics Postdoctoral Fellowship under award AST-2202388, and acknowledges previous support by the NSF under award 1847938.
Pulsar research at UBC is supported by an NSERC Discovery Grant and by CIFAR.
S.R.T. acknowledges support from an NSF CAREER award \#2146016.
C.U. acknowledges support from BGU (Kreitman fellowship), and the Council for Higher Education and Israel Academy of Sciences and Humanities (Excellence fellowship).
C.A.W. acknowledges support from CIERA, the Adler Planetarium, and the Brinson Foundation through a CIERA-Adler postdoctoral fellowship.
O.Y. is supported by the National Science Foundation Graduate Research Fellowship under Grant No. DGE-2139292.
A.G. acknowledges the support of the Department of Atomic Energy, Government of India, under project identification \# RTI 4002.
We thank the anonymous referee for their comments and suggestions.

\bibliographystyle{aasjournal}
\bibliography{ng12.5-3c66b-ecc}{}



\end{document}